\documentclass[aps,pre,reprint,superscriptaddress]{revtex4-1}

\usepackage{amsmath,amssymb}
\usepackage{graphicx}
\usepackage[italicdiff]{physics}
\usepackage{siunitx}
\usepackage{xparse}
\NewDocumentCommand\drv{m m}{f_{k_{#1}k_{#2}}}
\NewDocumentCommand\ddrv{m m}{\kappa_{k_{#1}k_{#2}}}
\usepackage{hyperref}

\begin{document}
\title{Vibrational properties of two-dimensional dimer packings near the jamming transition}

\author{Kumpei Shiraishi}
\email{kumpeishiraishi@g.ecc.u-tokyo.ac.jp}
\affiliation{Graduate School of Arts and Sciences, University of Tokyo, Komaba, Tokyo 153-8902, Japan}
\author{Hideyuki Mizuno}
\affiliation{Graduate School of Arts and Sciences, University of Tokyo, Komaba, Tokyo 153-8902, Japan}
\author{Atsushi Ikeda}
\affiliation{Graduate School of Arts and Sciences, University of Tokyo, Komaba, Tokyo 153-8902, Japan}
\affiliation{Research Center for Complex Systems Biology, Universal Biology Institute, University of Tokyo, Komaba, Tokyo 153-8902, Japan}

\date{\today}

\begin{abstract}
Jammed particulate systems composed of various shapes of particles undergo the jamming transition as they are compressed or decompressed.
To date, sphere packings have been extensively studied in many previous works, where isostaticity at the transition and scaling laws with the pressure of various quantities, including the contact number and the vibrational density of states, have been established.
Additionally, much attention has been paid to nonspherical packings, and particularly recent work has made progress in understanding ellipsoidal packings.
In the present work, we study the dimer packings in two dimensions, which have been much less understood than systems of spheres and ellipsoids.
We first study the contact number of dimers near the jamming transition.
It turns out that packings of dimers have ``rotational rattlers'', each of which still has a free rotational motion.
After correcting this effect, we show that dimers become isostatic at the jamming, and the excess contact number obeys the same critical law and finite size scaling law as those of spheres.
We next study the vibrational properties of dimers near the transition.
We find that the vibrational density of states of dimers exhibits two characteristic plateaus that are separated by a peak.
The high-frequency plateau is dominated by the translational degree of freedom, while the low-frequency plateau is dominated by the rotational degree of freedom.
We establish the critical scaling laws of the characteristic frequencies of the plateaus and the peak near the transition.
In addition, we present detailed characterizations of the real space displacement fields of vibrational modes in the translational and rotational plateaus.
\end{abstract}

\maketitle

\section{Introduction}\label{sec:intro}
Jammed particulate systems composed of (frictionless) particles are known to undergo the jamming transition when they are compressed or decompressed~\cite{van_Hecke_2010}.
In particular, the system of spherical particles has been studied extensively by many previous works~(see, e.g.,~Ref.~\cite{O_Hern_2003}).
At the transition point where the pressure reaches zero, the system becomes isostatic. 
The isostaticity here means that the number of constraints of the system equals the number of degrees of freedom~(DOF); that is, for sphere packings, the contact number equals $z_\text{iso} = 2d$ where $d$ is the spatial dimension.
Close to the transition, the excess contact number, $\Delta z \equiv z - z_\text{iso}$, scales as the square root of pressure $p$ or excess density $\Delta \phi \equiv \phi - \phi_J (> 0)$: $\Delta z \propto p^{1/2} \propto \Delta\phi^{1/2}$.
Mechanical properties such as elastic moduli also show the power-law scaling as a function of $p$ or $\Delta \phi$.
Moreover, soft modes emerge associated with the jamming transition. 
The vibrational density of states (vDOS) shows the characteristic plateau, which extends towards the zero frequency at the transition point~\cite{Silbert_2005}.
The onset frequency of the plateau, which is denoted by $\omega^{*}$, scales linearly with the excess contact number: $\omega^{*} \propto \Delta z$.
The variational argument~\cite{Wyart_EPL_2005,Wyart_PRE_2005,Yan_2016} and effective medium theory~\cite{Wyart_2010,DeGiuli_2014} have been proposed to explain this scaling behavior of vDOS.

Jammed systems composed of various shapes of nonspherical particles, such as ellipsoidal particles~\cite{Donev_2004,Donev_Stillinger_2004,Donev_2007,Mailman_2009,Zeravcic_2009,Schreck_2010,Delaney_2010,Xia_2014,Xu_Sun_2015,Schaller_2015,Yuan_2018}, sphero-cylinders~\cite{Williams_2003,Wouterse_2009,Zhao_2012,Meng_2016,Marschall_2018}, and composites of spherical particles~\cite{Baule_2013,Gaines_2016,Gaines_2017,Griffith_2018}, have also been studied numerically, theoretically, and experimentally.
One of the most studied nonspherical particles is the ellipsoidal particle.
It is found for ellipsoidal packings that the contact number becomes smaller than the isostatic value $z_\text{iso} = 2d_f$ ($d_f$ is the number of DOFs that an ellipsoidal particle possesses) at the jamming transition, which is referred to as hypostaticity~\cite{Donev_2004,Donev_2007,Mailman_2009,Zeravcic_2009,Schreck_2010,Schreck_2012}.
This hypostaticity has attracted attentions of previous works because it drastically alters the mechanical and vibrational properties near the jamming transition~\cite{Mailman_2009,Schreck_2012,Zeravcic_2009,Yunker_2011,Brito_2018}.
One might expect that the packings of ellipsoidal particles are not mechanically stable since the hypostaticity implies the shortage of constraints.
However, they actually have positive elastic moduli and do not have zero-frequency modes in the jammed phase $\phi > \phi_J$.
The hypostaticity indeed induces the floppy modes, which, however, are not exactly zero-frequency modes but are stabilized to become finite frequency modes by the repulsive interactions between ellipsoidal particles~\cite{Mailman_2009,Schreck_2012}.
Recent theoretical work~\cite{Brito_2018} proposed a variational argument to describe the vibrational states in ellipsoidal packings and explain the scaling behavior of the vDOS, including that of these floppy modes, near the jamming transition.
Here, we remark that ellipsoidal shape is not the only nonspherical shape to generate hypostatic packings.
Sphero-cylinders also show hypostaticity in their contact numbers, and vibrational properties were reported to be similar to those of ellipsoidal packings~\cite{VanderWerf_2018,Marschall_2018}.

In this work, we focus on another kind of nonspherical particles, dimer particles.
Compared to the ellipsoidal particles described above, dimer particles have been still much less understood.
The ellipsoidal particles show two major differences from the spherical particles.
One is asphericity of constituent particles, and the other is hypostaticity.
As a result, the vDOS of the ellipsoids has a number of different characteristics compared to spheres~\cite{Mailman_2009,Schreck_2012,Brito_2018}.
However, the dimer particles are known to be isostatic at the jamming transition~\cite{Schreck_2010}.
Namely, the dimer particles possess asphericity, but they are not hypostatic in the jammed phase.
In recent experimental and numerical works~\cite{Han_2012,Han_2013,Han_2015}, structures of two-dimensional asymmetric dimer packings were studied.
For vibrational properties, it has been reported that the dimer particles exhibit rotational vibrations in the low-frequency region below the translational band~\cite{Schreck_2010}.
However, the vDOS of the dimers has not been studied yet.
As a result, it remains to be addressed whether the vDOS of the dimers exhibits the critical scaling behaviors near the jamming transition, such as the extension of the characteristic plateau, which is indeed our main topic in this work.

This article is organized as follows.
In Section~\ref{sec:methods}, we describe the numerical protocol to generate dimer packings and also formulate the equation of motion and the dynamical matrix to study vibrational eigenmodes.
Section~\ref{sec:results} contains our numerical results for the dimer packings.
We start from the numerical results on the jamming density.
We next clarify that the present model has a limitation when the density is much higher than the jamming density.
Then, we study the contact numbers of dimer packings near the jamming density.
It turns out that when we analyze the contact numbers, we have to take care of the ``rotational rattlers'', which are a new type of rattler.
We finally present the results of vibrational properties of dimer systems, such as the vDOS, the contribution of rotational vibrations, and degree of localization.
Importantly, we find two characteristic plateaus in the vDOS: the translational plateau at higher frequency and the rotational plateau at lower frequency.
Accordingly, we define two characteristic frequencies: $\omega^*$ for the onset of the rotational plateau and $\omega_R$ for the rotational peak that separates the translational and rotational plateaus.
We numerically establish the scaling behaviors of $\omega^*$ and $\omega_R$ near the jamming transition with the excess density $\Delta\phi$ and with the aspect ratio of dimer particles.
Section~\ref{sec:conclusion} summarizes our results and gives conclusions and remarks.

\section{Model and Methods}\label{sec:methods}
\subsection{System description}\label{sec:system_descr}
\begin{figure}[tb]
 \centering
 \includegraphics[width=.5\columnwidth]{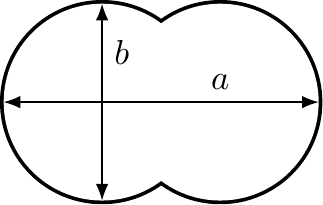}
 \caption{Conformation of dimer particle in 2D.
Aspect ratio is defined as $\alpha = a/b$.}
 \label{fig:dimer_shape}
\end{figure}

We generated a numerical system composed of dimer particles in two dimensions~(2D).
The number of constituent dimers $N$ ranges from \num{60} to \num{3600} in the present work.
Our model of a dimer particle is constructed by connecting two monomers that are modeled by harmonic disks.
Figure~\ref{fig:dimer_shape} illustrates conformation of the dimer particle.
We set the major axis of the dimer as $a$ and the minor axis as $b$, and define aspect ratio $\alpha$ as $\alpha = a/b$.
The system is a binary mixture in order to avoid crystallization: $2N/3$ of dimer particles have minor axis $b = \sigma$~(major axis $a = \sigma \alpha$), and $N/3$ have $b = 1.4 \sigma$~($a = 1.4 \sigma \alpha$), as in the previous work~\cite{Schreck_2010}, where the value of $\alpha$ is fixed to be identical for all dimer particles.
We also set the same mass for all dimer particles as $m$.

In our dimer model, each monomer of a dimer interacts with monomers belonging to the other dimers via purely repulsive harmonic potential~\cite{Schreck_2010}.
The total potential of the system is described as the summation of harmonic potential over all the monomers:
\begin{align}
V = \sum_{i>j} \sum_{n_i n_j} \frac{\epsilon}{2} \pqty{1 - \frac{r_{k_ik_j}}{\sigma_{ij}}}^2 H\pqty{1 - \frac{r_{k_ik_j}}{\sigma_{ij}}}, \label{potential}
\end{align}
where $i$ and $j$ represent the dimer $i$ and dimer $j$, respectively.
$k_i$, which represents a monomer of the dimer $i$, is composed of a set of integers: $k_i \equiv \pqty{i, n_i}$, where $n_i = \pm 1$ designates the monomer that belongs to the dimer $i$.
$\epsilon$ sets the energy scale, $r_{k_ik_j}$ is the distance between the monomer $k_i$ and monomer $k_j$,  $\sigma_{ij} = \pqty{b_i + b_j}/2$~($b_i$ and $b_j$ are the minor axes of dimers $i$ and $j$, i.e., the diameters of monomers $k_i$ and $k_j$, respectively), and $H\pqty{x}$ is the Heaviside step function: $H\pqty{x} = 1$ for $x \geq 0$ and $H\pqty{x} = 0$ for $x < 0$.
Throughout this paper, we use $m$, $\sigma$, and $\epsilon$ as units of mass, length, and energy, respectively.
We note that frequency is measured by $\tau = \pqty{m \sigma^2/\epsilon}^{-1/2}$.

We first prepared a random configuration at sufficiently small packing fraction $\phi_0 = \num{0.2}$ and minimized the potential energy using the FIRE algorithm~\cite{Bitzek_2006}.
Here, the periodic boundary conditions are implemented on the system.
The packing fraction $\phi$ of 2D dimers is defined as~\cite{Schreck_2010}
\begin{align}
 \phi = \frac{N_s b_s^2 + N_l b_l^2}{2L^2} \bqty{\pi - \cos^{-1}\pqty{\alpha-1} + \pqty{\alpha-1}\sqrt{\alpha\pqty{2-\alpha}}},
\end{align}
where $L$ is the system length, $N_s=2N/3$ ($N_l=N/3$) is the number of smaller (larger) particles, and $b_s=\sigma$ ($b_l=1.4\sigma$) is the length of the minor axis of smaller (larger) particles.
We then gradually compressed the system by increments of the packing fraction of $\delta\phi = \num{e-3}$.
After each compression, the energy minimization was executed.
Here, we note that the minimization is terminated when either of the following two conditions is satisfied~\cite{Gao_2006}:
(i) The total potential per particles is sufficiently small $V/N < \num{e-16}$; or 
(ii) The total potential energies of two successive FIRE steps $t$ and $t+1$ are nearly the same: $\abs{V_{t+1}-V_t}/V_t < \num{e-16}$.
We finally obtained the jamming transition point as the point where the potential energy per particle is in a range of $\num{1e-16} < V/N < \num{2e-16}$.
To numerically achieve this aim, we repeated compression and decompression on the system around the transition point.
Each time the compression/decompression is switched, $\delta\phi$ is multiplied by \num{0.5}.

Once we obtained the configuration at the transition point, we compressed the system to obtain the configurations as a function of the excess density $\Delta\phi = \phi - \phi_J$.
In these obtained packings, we recursively removed ``rattler'' particles that have contacts fewer than the number of DOF, $d_f \equiv 3$~(two translational DOFs and one rotational DOF).
In the following, the number of particles without rattler particles is denoted as $N$.

\subsection{Vibrational mode analysis}\label{sec:vib-analysis}
The Lagrangian $\mathcal{L}$ of the system is
\begin{align}
 \mathcal{L} = \sum_{i=1}^N \pqty{\frac{1}{2}{\dot{x}_i}^2 + \frac{1}{2}{\dot{y}_i}^2 + \frac{I_i}{2}{\dot{\theta}_i}^2} - V,
\end{align}
where $x_i$ and $y_i$ are spatial positions of the center of mass of the dimer $i$, and $\theta_i$ is rotational angle of the dimer $i$ with respect to the $x$-axis.
$V$ is the total potential energy defined by Eq.~\eqref{potential}, and $I_i$ is the inertia of the dimer $i$.
The detailed formulation of inertia is given in Appendix~\ref{app:moment_of_inertia}.
We employ the normal coordinate of dimer $i$ as $\pqty{x_i, y_i, \varphi_i}$ with $\varphi_i = \sqrt{I_i}\theta_i$ and that of whole the system as $\vb{r} = \pqty{x_1,y_1,\varphi_1,\dots,x_N,y_N,\varphi_N}$.
Then, the equation of motion~(Euler-Lagrange equation) is given as
\begin{align}
 \dv[2]{\vb{r}}{t} = -\pdv{V}{\vb{r}}. \label{eom}
\end{align}

When we suppose the harmonic vibrations around the equilibrium coordination $\vb{r}_0$, the equation of motion~\eqref{eom} is described as
\begin{align}
 \dv[2]{\vb{u}}{t} = -\mathcal{M} \vb{u}, \label{harmoniceom}
\end{align}
where $\vb{u} \equiv \vb{r}-\vb{r}_0$ is the displacement from $\vb{r}_0$.
$\mathcal{M}$ is the so-called dynamical matrix, and its elements are
\begin{align}
 \mathcal{M}_{kl} = \pdv[2]{V}{u_k}{u_l}, \label{dynmat}
\end{align}
where $k,l = 1,\dots,d_f N$, and $u_k$ is the $k$th component of $\vb{u}$.
Explicit formulations of $\mathcal{M}_{kl}$ are given in Appendix~\ref{app:dynmat}.
By diagonalizing the dynamical matrix, we obtained a set of eigenvalues, $\lambda^k$, and eigenvectors, $\vb{e}^k \equiv \pmqty{\vb{e}^k_1 & \cdots & \vb{e}^k_N}$, with $\vb{e}^k_i = \pmqty{e^k_{i,x} & e^k_{i,y} & e^k_{i,\varphi}}$~($i=1, \dots, N$).
The eigenfrequencies are given as $\omega^k = \sqrt{\lambda^k}$, and the eigenvectors are orthonormalized as $\vb{e}^k \cdot \vb{e}^l \equiv \sum_{i=1}^N \vb{e}^k_i \cdot \vb{e}^l_i = \delta_{kl}$ where $\delta_{kl}$ is the Kronecker delta.

In the present work, we performed vibrational eigenmode analysis on the ``unstressed system''.
Since the monomers are modeled by harmonic disks in the present model, forces acting between the monomers are always repulsive.
For this reason, we refer to this original state as the stressed system.
In the unstressed system, we retain all the stiffness~(i.e., the second derivative of potential) between monomers but drop the force in the analysis.
For the case of sphere packings, the theoretical understanding was first constructed based on the unstressed system~\cite{Wyart_EPL_2005,Wyart_2010}, which was then extended to the stressed system by considering effects of the forces~\cite{Wyart_PRE_2005,DeGiuli_2014}.
The unstressed system of sphere packings exhibits the characteristic plateau in vDOS and the critical behavior of this plateau near the jamming transition.
The forces in sphere packings do not affect this plateau and its critical behavior, whereas they make the system mechanically unstable~\cite{Lerner_2014} and alter the nature of the very low-frequency vibrational modes~\cite{Mizuno_2017}.
In particular, quasilocalized vibrational modes are induced by the repulsive forces~\cite{Mizuno_2017,Lerner_2018,Shimada_2018}.
In the present work, we limit ourselves to analyzing the vibrational properties of unstressed dimer packings.
Based on the results of unstressed systems, we will discuss the role of repulsive forces in dimer packings in the near future.

\section{Results}\label{sec:results}

\subsection{Jamming density}
\begin{figure}[tb]
 \centering
 \includegraphics[width=.99\columnwidth]{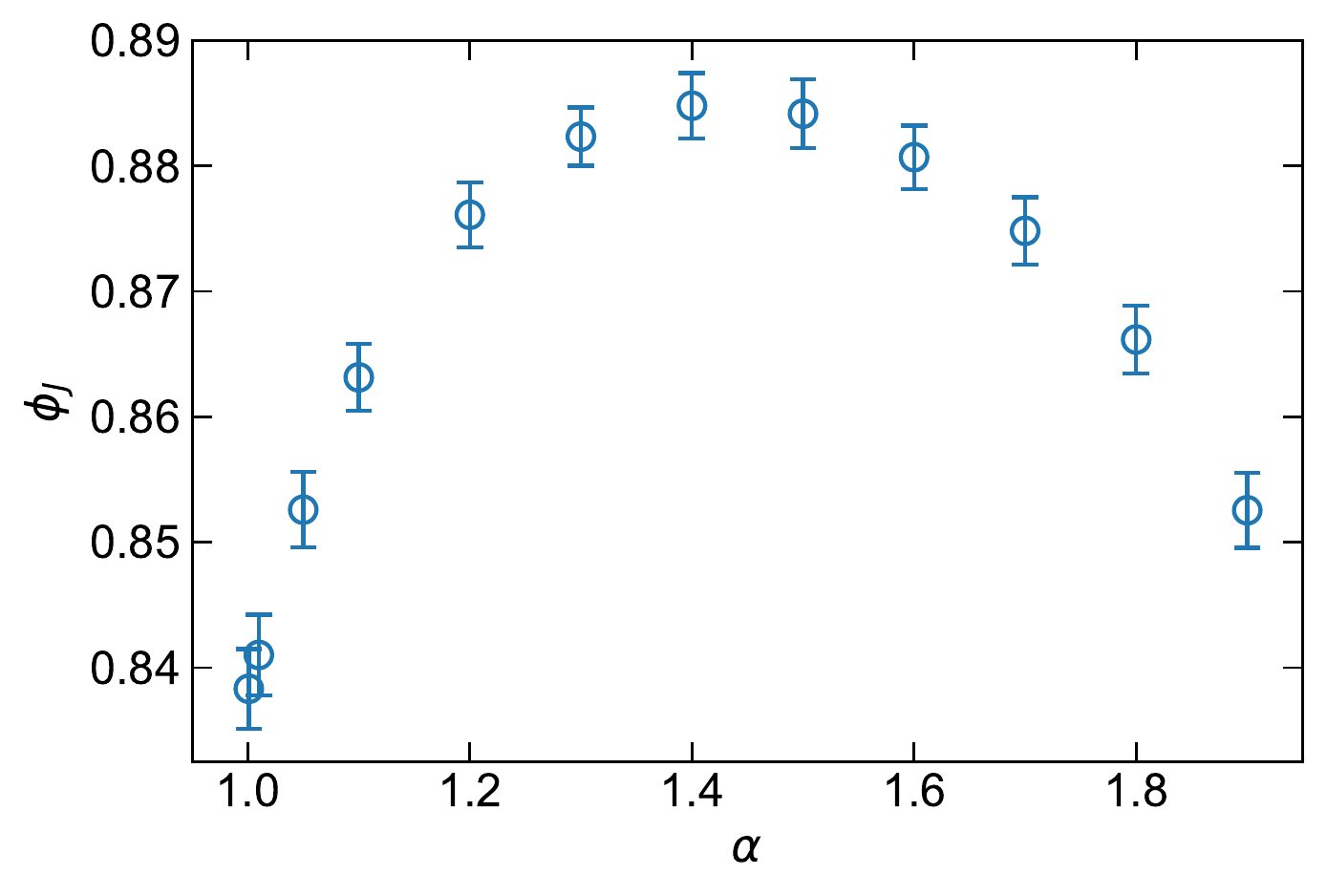}
 \caption{Packing fraction at the jamming transition $\phi_J$.
We plot $\phi_J$ as a function of the aspect ratio $\alpha$ that ranges from \num{1.001} to \num{1.9}.
The presented data are obtained by averaging over \num{50} to \num{1500} samples, and error bars are also determined by these samples.}
 \label{fig:phiJ}
\end{figure}

By following the procedure described in Section~\ref{sec:system_descr}, we generated the configurations at the jamming transition where the potential per particle $V/N$ is in $\num{1e-16} < V/N < \num{2e-16}$.
Figure~\ref{fig:phiJ} plots the jamming density $\phi_J$ as a function of the aspect ratio $\alpha$.
When $\alpha \to 1$, the jamming density approaches that for the binary mixtures of disks, $\phi_J \approx \num{0.84}$~\cite{O_Hern_2003}.
We also observe the nonmonotonic dependence of $\phi_J$ on $\alpha$, where the peak value is taken at approximately $\alpha = \num{1.4}$.
These results are consistent with previous observation by Ref.~\cite{Schreck_2010}.
Starting from the configurations at the jamming transition, we compress the system to obtain the configurations in the jammed phase $\phi > \phi_J$ (the excess density $\Delta \phi >0$).
We stress that we obtained $\phi_J$ for each configuration; then, $\Delta\phi$ was defined for each configuration separately.

\subsection{Limitations of the present model of dimers}\label{sec:model_limitations}
The rigorous way to simulate dimer particles is to consider the particle as one elastic element of the dimeric shape.
In this case, the potential energy of the system is given as the summation of the elastic deformation energies of all the dimer particles.
By contrast, the present model is an approximate model, where the potential energy is evaluated as the summation of the interparticle potentials between constituent monomers.
This approach introduces some limitations of the present model, which we now clarify.

\begin{figure}[tb]
 \centering
 \includegraphics[width=.99\columnwidth]{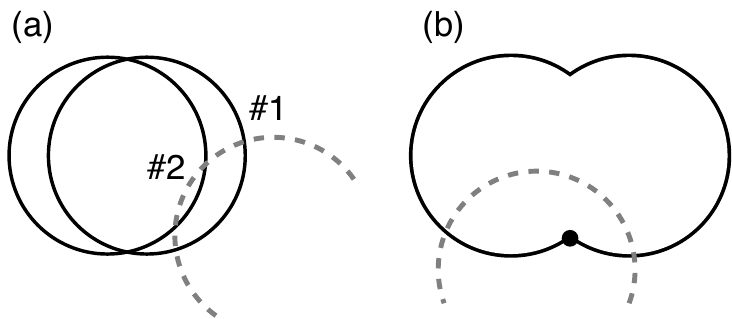}
 \caption{Two types of contacts, (a) double contact and (b) cusp contact, that cannot be addressed by the present modeling of dimers.
The arc described by the dashed line and the arc described by the solid line represent a monomer of a dimer and another dimer, respectively.
(a) Double contact: the dashed-line monomer of the dimer is embedded into the solid-line dimer so deeply that it interacts with both monomers of the solid-line dimer via contacts \# 1 and \# 2.
(b) Cusp contact: the dashed-line monomer of the dimer contacts the solid-line dimer, covering the whole cusp line of this other dimer.}
 \label{fig:double_cusp_contacts}
\end{figure}

There are two types of contacts that cannot be properly addressed by the present model.
One is ``double'' contact, which is illustrated in Fig.~\ref{fig:double_cusp_contacts} (a).
When a monomer of a dimer deeply overlaps with another dimer, the monomer makes contact with both constituent monomers of the other dimer.
This is an artificial effect in the present model, and as such, the contact cannot be properly addressed.
The other is the ``cusp'' contact, as illustrated in Fig.~\ref{fig:double_cusp_contacts} (b).
In this contact, a monomer of a dimer contacts another dimer covering whole the cusp line of the other dimer.
The present model is reasonable only for contacts where the deformation of a dimer can be decomposed into the deformations of two monomers.
This ``cusp'' contact clearly violates this condition, and such contacts cannot be properly addressed.
Thus, as long as we employ the present model, we have to discard configurations that have any of the double and the cusp types of contacts.
These contacts can occur at the higher packing fraction, and therefore, this condition gives an upper limit of the excess volume fraction $\Delta\phi$ that can be studied in the present model.

\begin{figure}[tb]
 \centering
 \includegraphics[width=.99\columnwidth]{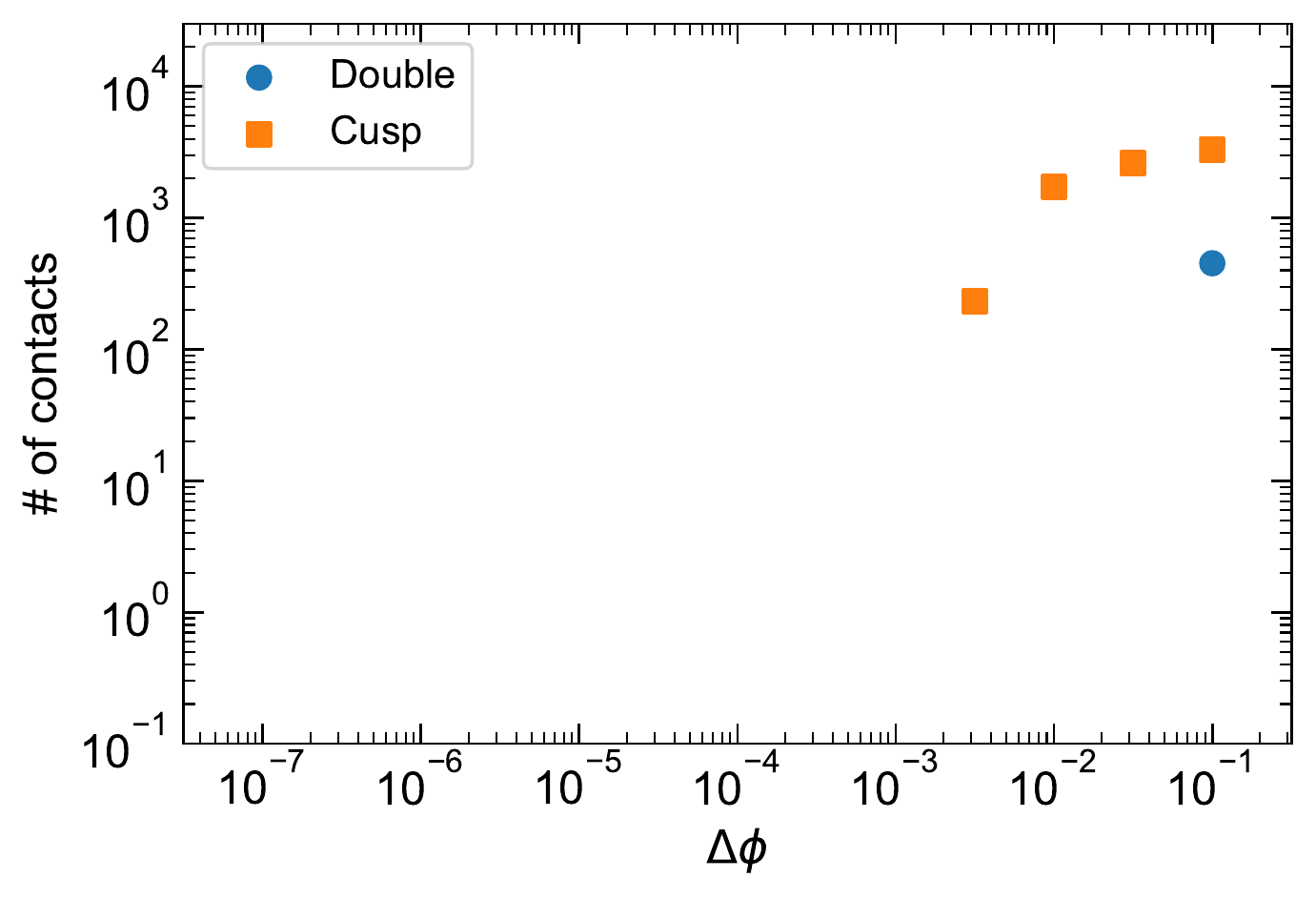}
 \caption{Plot of the number of double and cusp contacts as a function of the excess volume fraction $\Delta\phi$.
The aspect ratio is $\alpha = \num{1.1}$.
The system size is $N = \num{3600}$.
We obtain the presented values by averaging over \num{100} samples at each $\Delta\phi$.
For the definition of the double and the cusp contacts, refer to Fig.~\ref{fig:double_cusp_contacts} and the main text.}
 \label{fig:double-a1.1}
\end{figure}

We count the number of double and cusp contacts in configurations at each excess density $\Delta\phi$.
Figure~\ref{fig:double-a1.1} plots the number of these contacts as a function of $\Delta\phi$ for the case of aspect ratio $\alpha = \num{1.1}$.
The system size is $N = \num{3600}$.
The number of cusp contacts is exactly zero up to $\Delta\phi = \num{e-3}$ and then becomes finite for $\Delta\phi \geq \num{e-2.5}$.
The number of the double contacts is exactly zero up to $\Delta\phi = \num{e-1.5}$.
Thus, configurations with $\Delta\phi \leq \num{e-3}$ include neither of these two contacts, and the upper limit of the excess density is determined as $\Delta\phi = \num{e-3}$ for $\alpha = \num{1.1}$.
We perform the same analysis for several different aspect ratios and obtain the upper limits of the excess densities, as $\Delta\phi = \num{e-1.5}$ for $\alpha = \num{1.5}$, and $\Delta\phi = \num{e-4.5}$ for $\alpha = \num{1.03}$.
We find that these values of the upper limits are independent of the system size.
In the present work, we limit ourselves to the densities below these upper limits.

\subsection{Contact number}\label{sec:contact}
\begin{figure}[tb]
 \centering
 \includegraphics[width=.99\columnwidth]{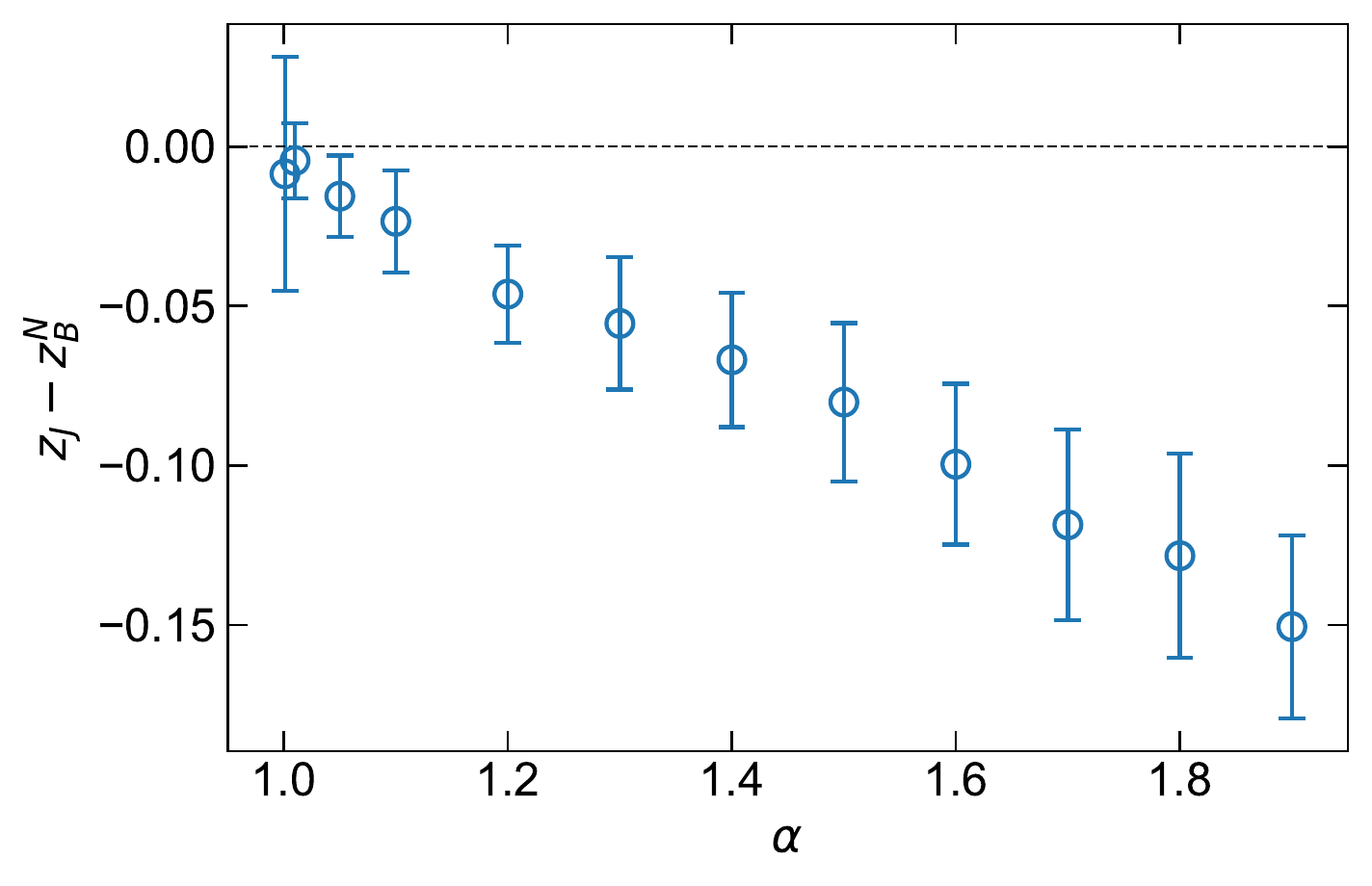}
 \caption{Contact number $z_J$ at the jamming transition $\phi_J$.
We plot $z_J$ as a function of aspect ratio $\alpha$.
The system size is $N = \num{240}$.
The horizontal line indicates zero.
The presented data are obtained by averaging over \num{50} to \num{1500} samples.}
 \label{fig:contact_shortage}
\end{figure}

We next study the contact number $z_J$ at the jamming transition $\phi_J$.
We first review the known results for $z_J$ of spherical particles.
Generally, when each particle has $d_f$ DOFs, the isostatic condition in the infinite system size is $z_\text{iso} = 2d_f$.
For spheres, the number of DOFs equals the spatial dimension, $d_f = d$; then, $z_\text{iso} = 2d$.
However for finite-size systems with $N$ particles, two types of finite-size effects appear~\cite{Goodrich_2012}.
First, because the systems are allowed to have $d$ global translations, the isostatic contact number is reduced to $z^N_\text{iso} = z_\text{iso} - 2d/N$.
Second, because the systems must have positive bulk modulus, the isostatic contact number is increased to $z^N_B = z^N_\text{iso} + 2/N = 2d - 2d/N + 2/N$.
It is established for spheres that $z_J = z^N_B$ holds for any system size~\cite{Goodrich_2012}.
Note that both $z^N_\text{iso}$ and $z^N_B$ converge to $z_\text{iso}$ in the thermodynamic limit of $N \to \infty$.

We now discuss whether this argument holds for the dimer packings in 2D.
Following the above arguments, we obtain the isostatic contact numbers for dimers:
\begin{align}
z^N_\text{iso} &\equiv 2d_f - \frac{2d}{N}, \label{naive-zNiso} \\
z^N_B &\equiv z^N_\text{iso} + \frac{2}{N} = 2d_f - \frac{2d}{N} + \frac{2}{N}, \label{naive-zNB}
\end{align}
with the DOFs $d_f = 3$ and the spatial dimension $d = 2$ in our 2D dimer system.
To test the relation of $z_J = z^N_B$ for dimers, we numerically measured $z_J$ of dimers for various aspect ratios and plot the difference $z_J$ from $z_B^N$ of Eq.~\eqref{naive-zNB} in Fig.~\ref{fig:contact_shortage}.
Clearly, $z_J$ takes a smaller value than $z^N_B$, and the difference grows larger as the aspect ratio increases.

\begin{figure}[tb]
 \centering
 \includegraphics[width=.99\columnwidth]{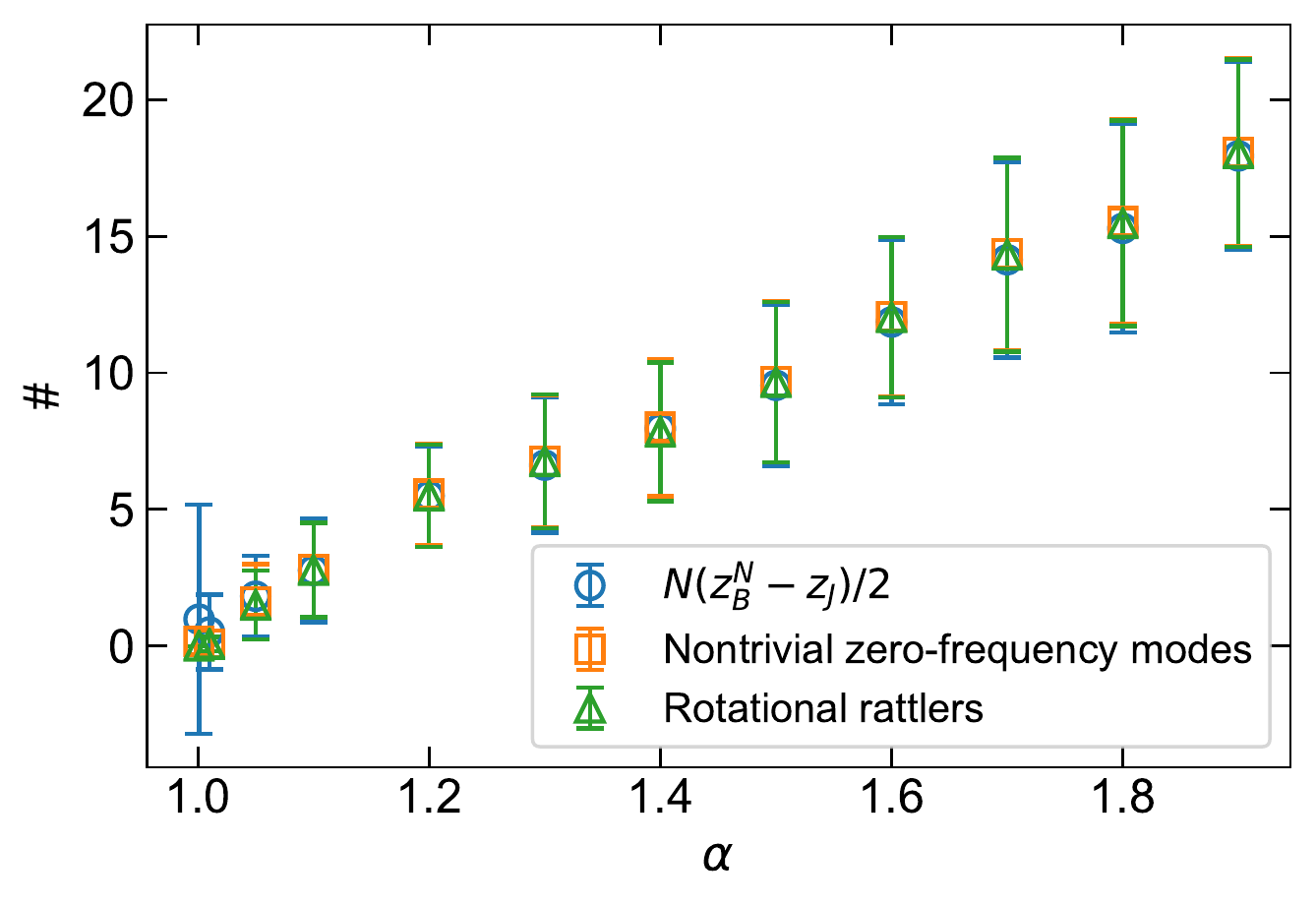}
 \caption{The numbers of shortages of contacts $N \pqty{z^N_B - z_J}/2$, nontrivial zero-frequency modes~(excluding trivial global translational modes), and rotational rattlers.
We plot these three numbers as functions of aspect ratio $\alpha$.
The system size is $N = \num{240}$.
The presented data are obtained by averaging over \num{50} to \num{1500} samples.
The definition of rotational ratters is given in Fig.~\ref{fig:zeromode_schematic} and the main text.}
 \label{fig:mathch_three_quantities}
\end{figure}

To understand the origin of $z_J < z^N_B$, we look into the vibrational eigenmodes obtained from the dynamical matrix~\footnote{Here, we study vibrational eigenmodes in the original stressed system. On the other hand, all of the results presented in Section~\ref{sec:vib_prop} are obtained from the unstressed system. Note that the stressed and unstressed system are identical exactly at the jamming density.}.
Remarkably, we find zero-frequency modes, in addition to two trivial zero-frequency modes of global translations.
Figure~\ref{fig:mathch_three_quantities} compares the number of these nontrivial zero-frequency modes~(squres) and the number of shortages of contacts, $N \pqty{z^N_B - z_J}/2$~(circles). 
These two numbers are the same for all the studied aspect ratios. 
Thus, the zero-frequency modes directly correspond to the shortage of $z_J$ from $z^N_B$. 

\begin{figure}[tb]
 \centering
 \includegraphics[width=.65\columnwidth]{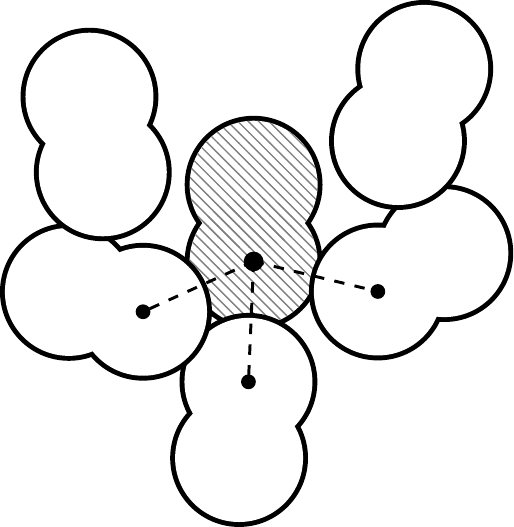}
 \caption{Schematic illustration of rotational rattlers.
The dimer of the oblique line pattern possesses three contacts, which is equal to the number of DOF, so it is not a rattler.
However, three contacts concentrate in only one monomer while the other monomer has no contacts.
In this situation, the dimer can rotate without any energy cost, i.e., it produces a zero-frequency vibrational mode.}
 \label{fig:zeromode_schematic}
\end{figure}

Then, we look into the nature of these zero-frequency modes.
We find that these modes are spatially localized to the dimers that have characteristic contacts as illustrated in Fig.~\ref{fig:zeromode_schematic}: they have more than three contacts, i.e., they are not rattler particles, but all of the contacts are concentrated in one monomer while the other monomer has no contacts.
These dimers can obviously rotate around one monomer without any energy cost; thus, they can produce zero-frequency modes~\cite{Alexander_1998}.
These dimers can be considered as another type of rattler particle: their translational DOFs are constrained by their three or more contacts, but the rotational DOFs are not.
In this paper, we refer to these dimer particles to as ``rotational rattlers''.
Figure~\ref{fig:mathch_three_quantities}~(triangles) shows that the number of rotational rattlers coincides with the number of nontrivial zero-frequency modes and the shortage of contacts for all the studied aspect ratios.
Thus, we conclude that the dimer packings have rotational rattlers that cause the nontrivial zero-frequency vibrational modes in addition to the apparent shortage of contacts~\footnote{The number of rotational rattlers, the number of nontrivial zero-frequency modes, and the shortage of contacts are exactly the same for approximately \SIrange[range-phrase=--]{80}{90}{\percent} of the configurations at $\phi_J$. They are almost the same but not exactly the same for the rest \SIrange[range-phrase=--]{10}{20}{\percent} of the configurations, for which we do not have clear explanation.}.

For the proper definition of the isostaticity of dimers, the rotational DOFs of rotational rattlers should be excluded, as in the case of the translational DOFs of usual rattlers.
This exclusion is achieved by replacing the total number of DOFs $N d_f$ with $N d_f - N^\prime$, where $N^\prime$ is the number of rotational rattlers.
This leads to the modified definition of the isostatic contact numbers for dimers:
\begin{align}
\tilde{z}^N_\text{iso} &\equiv 2d_f - \frac{2\pqty{N^\prime + d}}{N}, \label{tilde-zNiso} \\
\tilde{z}^N_B &\equiv \tilde{z}^N_\text{iso} + \frac{2}{N} = 2d_f - \frac{2\pqty{N^\prime + d}}{N} + \frac{2}{N}. \label{tilde-zNB}
\end{align}
Hereafter, we use these definitions as the isostatic contact numbers. 

\begin{figure}[tb]
 \centering
 \includegraphics[width=.99\columnwidth]{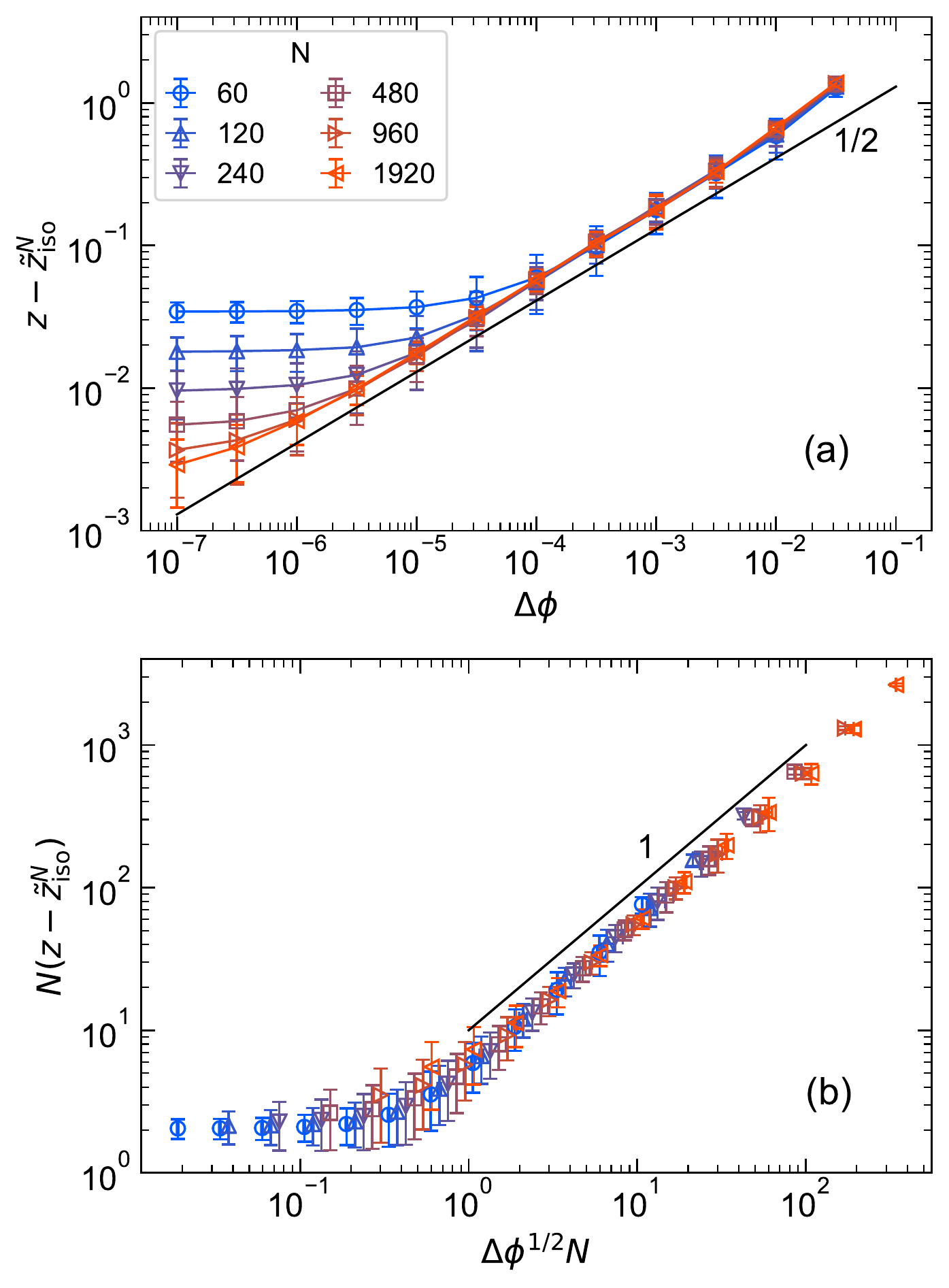}
 \caption{Finite-size scaling of the excess contact number.
(a) Plot of the excess contact number $\Delta z = z - \tilde{z}^N_\text{iso}$ as a function of excess density $\Delta\phi$.
(b) Plot of the scaled excess contact number $N\pqty{z - \tilde{z}^N_\text{iso}}$ as a function of the scaled excess density $\Delta\phi^{1/2} N$.
The aspect ratio is $\alpha = \num{1.5}$.
We observe the scaling law of $\Delta z \sim \Delta\phi^{1/2}$ in the high-density regime.
The presented data are obtained by averaging over \num{1000} to \num{5000} samples.}
 \label{fig:dzdp}
\end{figure}

Finally, we focus on the contact number at $\phi > \phi_J$.
Although we focus on $\alpha = \num{1.5}$ in the following, all the other aspect ratios exhibit similar behaviors.
In the case of spherical particles, the excess contact number $\Delta z = z - z_\text{iso}$ follows the square root law $\Delta z \sim \Delta\phi^{1/2}$~\cite{O_Hern_2003}.
More precisely, the excess contact number of $N$ spherical particles follows the finite-size scaling law $N\pqty{z - z^N_\text{iso}} = F\pqty{\Delta\phi^{1/2} N}$, where $F(x) \sim 1$ for small $x$ and $F(x) \sim x$ for large $x$~\cite{Goodrich_2012}~\footnote{In~\cite{Goodrich_2012}, the pressure $p$ is used instead of the excess packing fraction $\Delta\phi$.}.
The existence of the finite-size scaling law demonstrates that the jamming transition is a phase transition.
Here, we test whether these laws hold for the jamming transition of dimers or not.
Employing the value of $\tilde{z}^N_\text{iso}$ in Eq.~\eqref{tilde-zNiso}, we plot the excess contact number of dimers $z - \tilde{z}^N_\text{iso}$ as a function of the excess density $\Delta\phi$ for several different system sizes $N$ in Fig.~\ref{fig:dzdp} (a).
At the high packing fractions, the excess contact number follows the square root law $\Delta z \sim \Delta\phi^{1/2}$, as in the case of spheres~\cite{O_Hern_2003}.
We also observe that $\Delta z$ deviates from the square root law when approaching the jamming density, which is a signature of the finite-size effect.
We then replot data in the form of $N\pqty{z - \tilde{z}^N_\text{iso}}$ versus $\Delta\phi^{1/2} N$ in Fig.~\ref{fig:dzdp} (b).
Clearly, all the data points collapse onto a single master function $N \pqty{z - \tilde{z}^N_\text{iso}} = G\pqty{\Delta\phi^{1/2}N}$, where $G\pqty{x} \sim 1$ for small $x$ and $G\pqty{x} \sim x$ for large $x$.

Therefore, the results in this subsection establish that, when the rotational rattlers are properly excluded, the excess contact number of dimers near the jamming obeys the same critical law and finite-size scaling law as those of spheres.
These results also allow us to determine the range of densities in which the finite-size effect is negligible, that is, $\Delta\phi \geq \num{e-7}$ for $N = \num{3600}$ and $\Delta\phi \geq \num{e-6}$ for $N = \num{1200}$.
Note that this lower bound can be extended by increasing the system size, while the upper bound discussed in Section~\ref{sec:model_limitations} cannot.

\subsection{Vibrational properties}\label{sec:vib_prop}
In this subsection, we will discuss the nature of vibrational eigenstates, which are obtained by diagonalizing the dynamical matrices as explained in Section~\ref{sec:vib-analysis}.
We emphasize that the results are from the unstressed system.

To characterize the statistics of the eigenfrequencies $\omega^k$, we calculate the vDOS:
\begin{align}
 g\pqty{\omega} = \frac{1}{d_fN - N^\prime - d}\sum_{k=1}^{d_fN - N^\prime - d}\delta\pqty{\omega - \omega^k},
\end{align}
where $\delta\pqty{x}$ is the Dirac delta function.
We also analyze the nature of the eigenvectors $\vb{e}^k$ by quantifying the contribution of translational and rotational DOFs, and the participation ratio.
The contribution of translational DOFs to mode $k$ is defined as
\begin{align}
 T_k = \sum_{i=1}^N \bqty{\pqty{e^k_{i,x}}^2 + \pqty{e^k_{i,y}}^2},
\end{align}
as in the case of ellipsoids~\cite{Mailman_2009}.
The contribution of rotational DOFs is the counterpart of $T_k$, and defined as
\begin{align}
 R_k = 1 - T_k \label{Rk}.
\end{align}
The participation ratio $p_k$ is defined as
\begin{align}
 p_k = \frac{\pqty{\sum^N_{i=1}\abs{\vb{e}^k_i}^2}^2}{N\sum^N_{i=1}\abs{\vb{e}^k_i}^4}. \label{PR}
\end{align}
as in the case of spheres~\cite{Mazzacurati_1996}.
The participation ratio $p^k$ measures the extent of vibrational localization in the mode $k$.
We emphasize that $R_k$ and $p_k$ give the information of the eigenvectors in normal coordinates described in Section~\ref{sec:vib-analysis}.

\subsubsection{General nature of vibrational eigenmodes}
\begin{figure}[tb]
 \centering
 \includegraphics[width=.99\columnwidth]{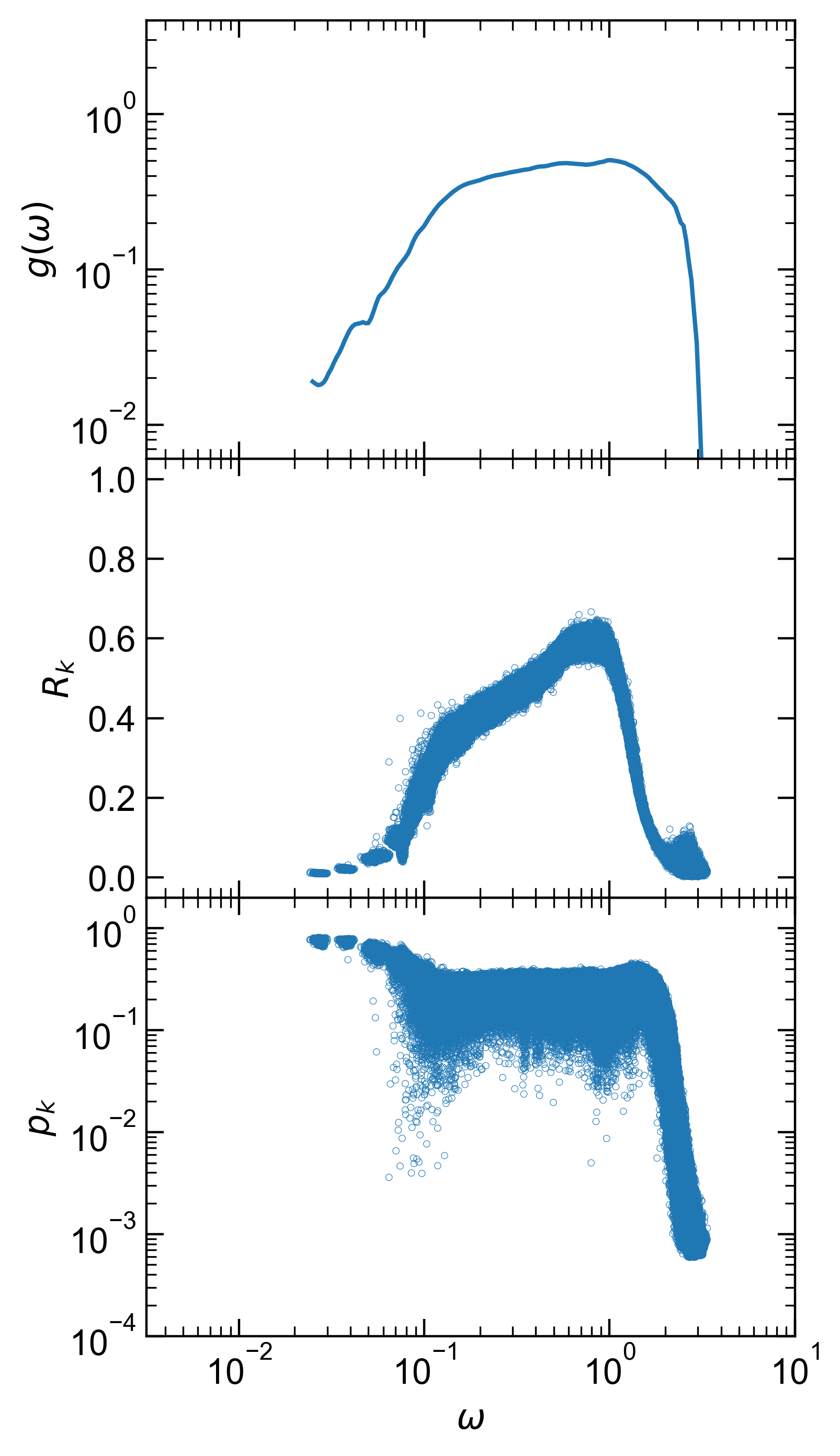}
 \caption{Vibrational eigenstates for aspect ratio $\alpha = \num{1.5}$.
We plot the vDOS $g\pqty{\omega}$, contribution of rotational DOFs $R_k$, and the participation ratios $p_k$.
The excess volume fraction is $\Delta\phi = \num{e-2}$, and the system size is $N = \num{3600}$.
The values of $g\pqty{\omega}$ are obtained by averaging over \num{100} samples, and the other quantities are presented by the scatter plot of all the modes in all \num{100} samples.}
 \label{fig:DOS_Rk_PR_a1.5}
\end{figure}

\begin{figure}[tb]
 \centering
 \includegraphics[width=.99\columnwidth]{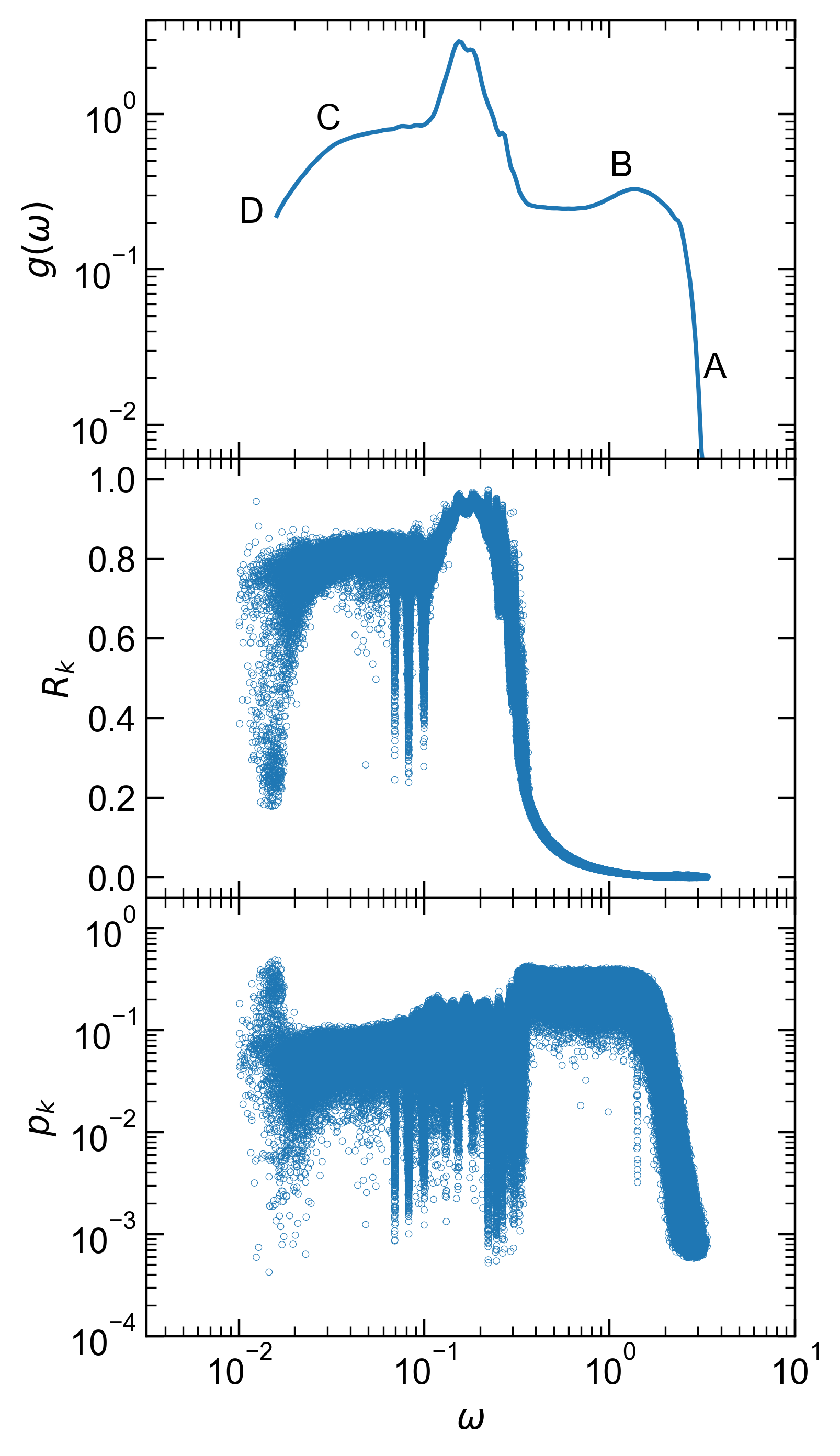}
 \caption{Vibrational eigenstates for aspect ratio $\alpha = \num{1.1}$.
We plot quantities of $g\pqty{\omega}$, $R_k$, $p_k$ as in Fig.~\ref{fig:DOS_Rk_PR_a1.5}.
The excess volume fraction is $\Delta\phi = \num{e-3}$, and the system size is $N = \num{3600}$.
Letters of ``A'' to ``D'' indicated in the panel of $g\pqty{\omega}$ correspond to the modes visualized in Fig.~\ref{fig:mode_visualize}.
See also the caption of Fig.~\ref{fig:DOS_Rk_PR_a1.5}.}
 \label{fig:DOS_Rk_PR_a1.1}
\end{figure}

We first focus on the vibrational states of dimers with the aspect ratio $\alpha = \num{1.5}$, which are presented in Fig.~\ref{fig:DOS_Rk_PR_a1.5}.
The excess packing fraction is $\Delta\phi = \num{e-2}$, and the system size is $N = \num{3600}$.
The top panel of Fig.~\ref{fig:DOS_Rk_PR_a1.5} shows the vDOS.
The structure of the vDOS is very similar to that of sphere packings.
In particular, it has the characteristic plateau at intermediate frequencies.
The second top panel shows the contribution of rotational DOFs, $R_k$.
We see that $R_k \approx \num{0}$ for the high-frequency edge and low-frequency regime below the plateau.
On the other hand, $R_k$ takes approximately \num{0.3} to \num{0.6} in the intermediate, plateau regime.
Therefore, the rotational DOFs are relevant only in the vibrational states in the plateau region.
The third top panel shows the participation ratio, $p_k$.
The plot demonstrates that the modes in the high-frequency edge are strongly localized and the modes in the plateau regime are extended, which are very similar to the case for spheres.
We also observe that the modes in the low-frequency regime below the plateau are largely extended and their eigenfrequencies are distributed discretely, which suggests that these modes are phonons.
We indeed confirm that the frequencies of these modes are consistent with the discretized levels of acoustic phonons of this system.
Therefore, at $\alpha = \num{1.5}$, the vibrational states of dimers are composed of the localized high-frequency edge, the intermediate plateau, and low-frequency acoustic phonons.
These features are very similar to those of spheres, apart from the fact that the rotational DOFs are relevant in the plateau modes for dimers.

We turn our attention to the vibrational states of dimers with the aspect ratio $\alpha = \num{1.1}$.
The excess packing fraction is $\Delta\phi = \num{e-3}$, and the system size is $N = \num{3600}$.
The vDOS, $R_k$, and $p_k$ are presented in Fig.~\ref{fig:DOS_Rk_PR_a1.1}, in the same way as in Fig.~\ref{fig:DOS_Rk_PR_a1.5}.
The vDOS for $\alpha = \num{1.1}$ appears to be very different from that for $\alpha = \num{1.5}$; we observe two characteristic plateaus, and they are separated by a peak.
$R_k \approx \num{0}$ in the plateau of high frequency, while $R_k \approx \num{0.8}$ in that of low frequency.
This result indicates that the rotational DOFs do not play any role in the vibrations in the high-frequency plateau, while they are relevant in the low-frequency plateau.
Hereafter, we refer to the high-frequency plateau as the ``translational plateau'' and the low-frequency plateau as the ``rotational plateau''.
The participation ratio of the translational plateau modes is $p_k \approx \num{0.3}$, while the rotational plateau modes take smaller values of $p_k \lesssim \num{0.1}$.
This observation suggests that the rotational plateau modes can be localized in the space of normal coordinates (we will discuss this point based on the real space displacement field later in Section~\ref{sec:plateaumode}).
Interestingly, $R_k$ takes the largest values in the peak at approximately $\omega \approx \num{0.2}$.
We refer to this peak as the ``rotational peak''.
In the regime of $\num{0.06} \lesssim \omega \lesssim \num{0.3}$ including this rotational peak, the distribution of eigenfrequencies are discretized.
We speculate that these discretized modes are optical types of modes.
To fully clarify the nature of these modes, however, we need to analyze the dynamic structure factors~\cite{Wang_2015}, which is beyond the scope of this work.
In summary, the vibrational states of dimers at $\alpha = \num{1.1}$ are characterized by the emergence of the translational and rotational plateaus that are separated by the rotational peak.

We repeated a similar analysis for various different aspect ratios ($\num{1.001} \leq \alpha \leq \num{1.9}$).
We find that the results of larger aspect ratios are similar to that for $\alpha = \num{1.5}$, while those for smaller aspect ratios are similar to $\alpha = \num{1.1}$.
Therefore, the vibrational states of dimers are categorized into these two types of behaviors, depending on the aspect ratio.

\subsubsection{Critical behaviors near the jamming transition}

\begin{figure}[tb]
 \centering
 \includegraphics[width=.99\columnwidth]{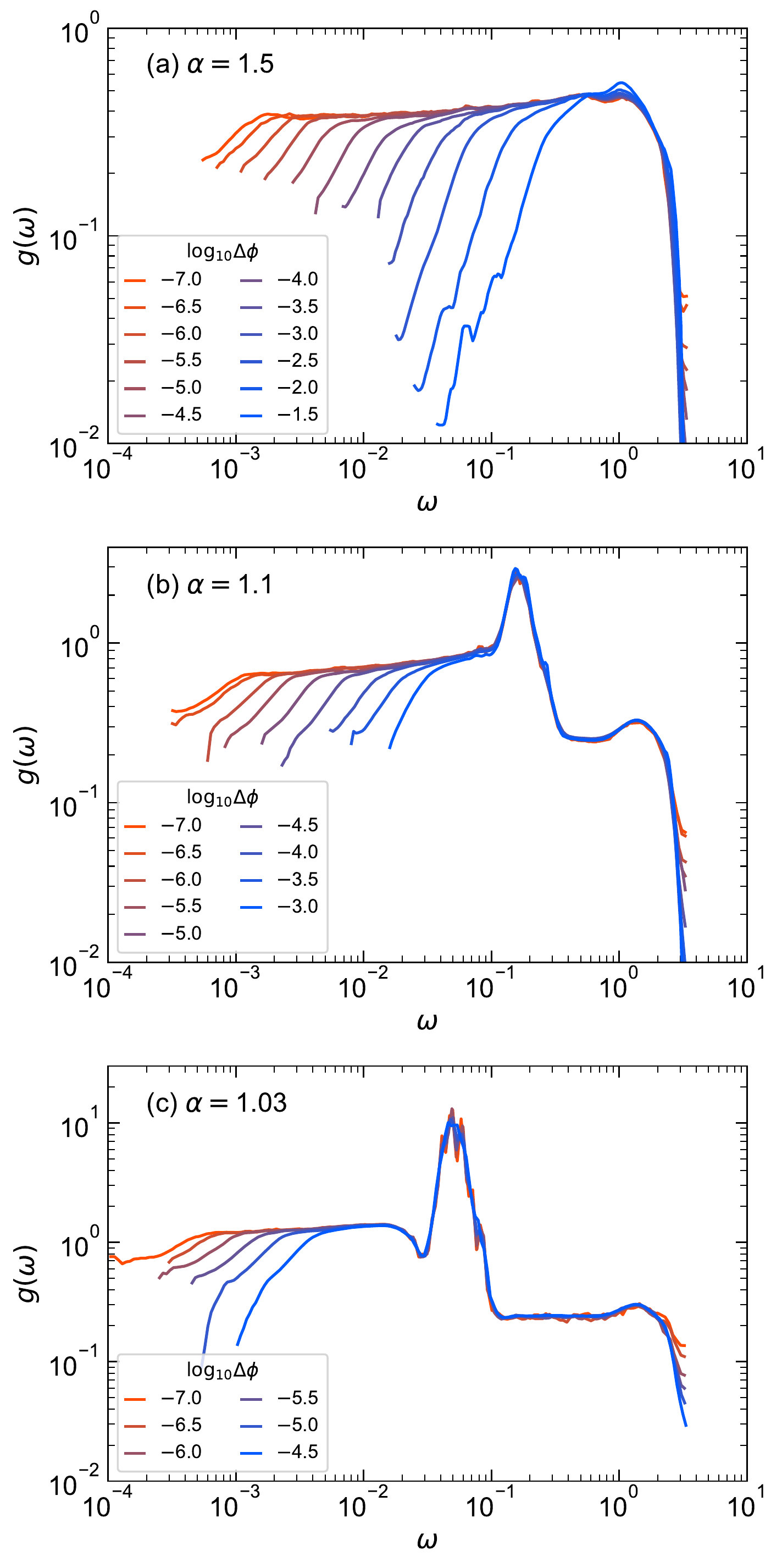}
 \caption{Packing fraction dependence of the vibrational density of the state for aspect ratio (a) $\alpha = \num{1.5}$, (b) $\alpha = \num{1.1}$, and (c) $\alpha = \num{1.03}$.
The system size is $N = \num{3600}$.
The presented data are obtained by averaging over \num{100} samples.}
 \label{fig:DOS}
\end{figure}

We now study the packing fraction dependence of vDOS near the jamming transition.
Figure~\ref{fig:DOS} plots the vDOS $g\pqty{\omega}$ for several different values of packing fraction $\Delta\phi$ and for aspect ratio $\alpha = \num{1.5}$ in (a), $\alpha = \num{1.1}$ in (b), and $\alpha = \num{1.03}$ in (c).
For (a) $\alpha = \num{1.5}$, the vDOS shows the characteristic plateau at intermediate frequencies.
We recall that the vibrational states in the plateau have contributions from rotational DOFs.
As we approach the jamming transition as $\Delta\phi \to 0$, this plateau extends towards zero frequency, which is similar to that observed in the vDOS of spheres.
For (b) $\alpha = \num{1.1}$ and (c) $\alpha = \num{1.03}$, the vDOSs have the translational plateau at high frequency and the rotational plateau at low frequency, which are separated by the rotational peak.
As we approach the transition, the rotational plateau extends towards zero frequency, while the translational plateau and the rotational peak are nearly unchanged.

\begin{figure}[tb]
 \centering
 \includegraphics[width=.99\columnwidth]{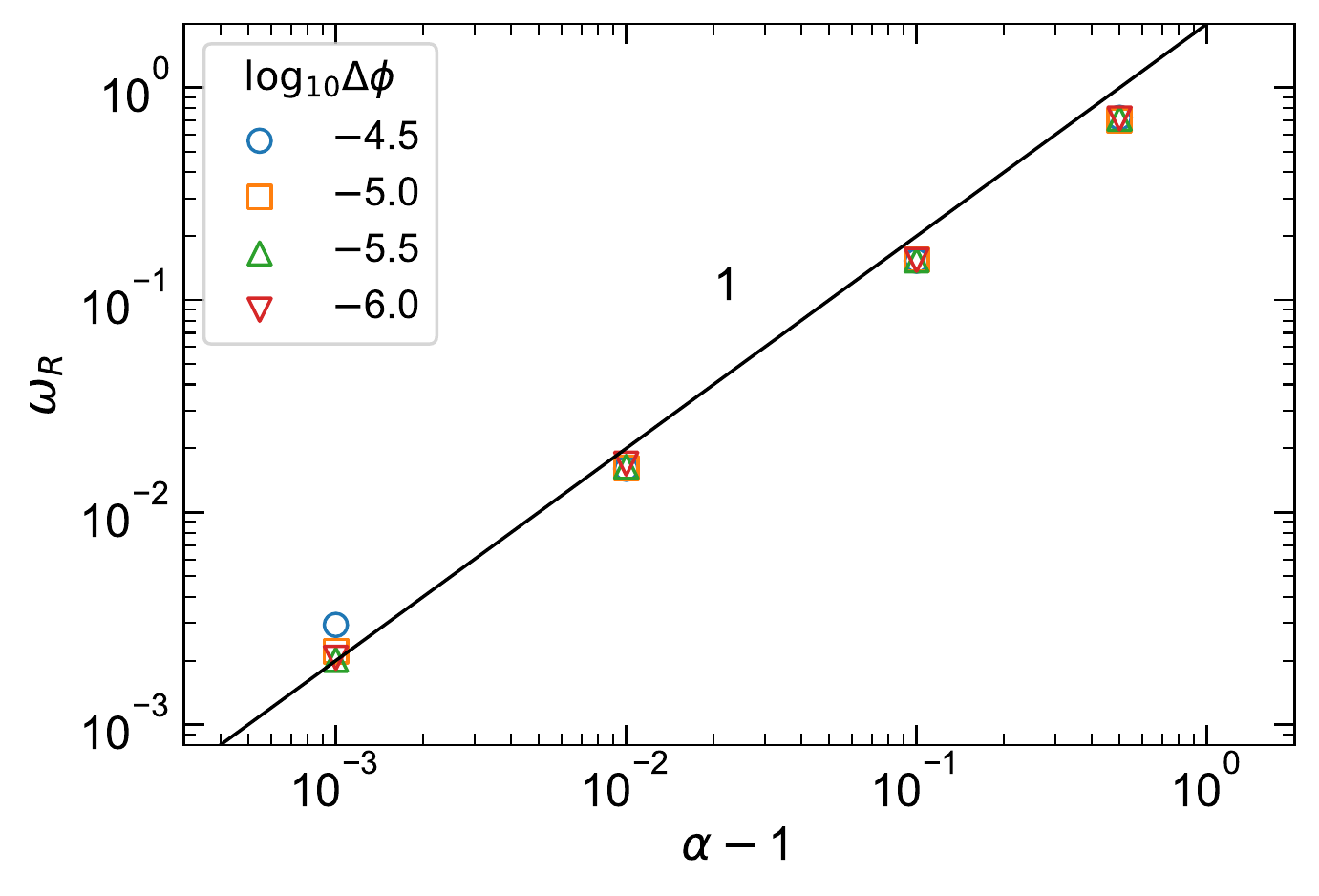}
 \caption{The frequency $\omega_R$ to characterize the rotational peak in the vDOS.
We plot $\omega_R$ as a function of the aspect ratio of $\alpha-1$ for several different volume fractions $\Delta\phi$.
$\omega_R$ scales linearly with $\alpha-1$, as $\omega_R \propto \alpha - 1$.}
 \label{fig:omegaR-alpha}
\end{figure}

From the above observations, we define two characteristic frequencies: $\omega_R$ for the frequency of the rotational peak, and $\omega^{*}$ for the onset of the rotational plateau.
We first study the excess density $\Delta\phi$ and aspect ratio $\alpha$ dependencies of $\omega_R$.
We used $N = \num{1200}$ systems for this analysis.
Note that although the rotational peak is absent for $\alpha = \num{1.5}$, we still observe the sharp increase in $R_k$ as shown in Fig.~\ref{fig:DOS_Rk_PR_a1.5}.
Therefore as a reference, we report the frequency at which $R_k$ takes the maximum as $\omega_R$ for $\alpha = \num{1.5}$.
Figure~\ref{fig:omegaR-alpha} plots the $\omega_R$ against $\alpha - 1$ for various $\Delta\phi$.
$\omega_R$ is nearly independent from $\Delta\phi$ but increases with $\alpha$.
This increase can be fitted into the linear function of $\alpha - 1$, especially for the small $\alpha$.
This $\alpha$ dependence is natural for the fundamental frequency scale of rotation, as explained below.
When we consider purely rotational vibration, the equation of motion (Eq.~\eqref{harmoniceom}) reduces to
\begin{align}
\dv[2]{\varphi_l}{t} = -\sum_{k=1}^N \mathcal{M}_{\varphi_l,\varphi_k} \pqty{ \varphi_k - \varphi_{0k} }.
\end{align}
As shown in Eq.~\eqref{Mthetatheta} of Appendix~\ref{app:dynmat}, the dynamical matrix elements $\mathcal{M}_{\varphi_l,\varphi_k}$ are proportional to $\pqty{\alpha-1}^2$; therefore, the fundamental frequency scale of rotation is proportional to $\alpha-1$.
This $\alpha$ dependence originates from the geometric fact that the displacement of monomers due to the pure rotation of dimers is proportional to $\alpha-1$ and causes an energy cost proportional to $\pqty{\alpha-1}^2$ (with constant stiffness).
Therefore, $\omega_R \propto \pqty{\alpha -1}$ is reasonable geometrically.

In relation to the scaling behavior of $\omega_R \propto \pqty{\alpha - 1}$, we can discuss the heights of the rotational and translational plateaus of the vDOS.
Figures~\ref{fig:DOS} (b) and (c) show that the rotational plateau increases with decreasing $\alpha$, whereas the translational plateau decreases with decreasing $\alpha$.
These tendencies are reasonable as discussed below.
We establish that the rotational peak at $\omega_R$ separates the rotational and translational plateaus.
Suppose that most of the translational DOFs and the rotational DOFs are confined in the translational~($\omega > \omega_R$) and rotational~($\omega < \omega_R$) plateaus, respectively, which is reasonably supported by the values of $R_k$ in Fig.~\ref{fig:DOS_Rk_PR_a1.1}.
Then, when $\omega_R$ becomes close to the zero frequency as $\alpha$ approaches unity, the frequency regime in which rotational DOFs emerge decreases, and the height of the rotational plateau increases.
We can observe this behavior in Fig.~\ref{fig:DOS}.

\begin{figure}[tb]
 \centering
 \includegraphics[width=.99\columnwidth]{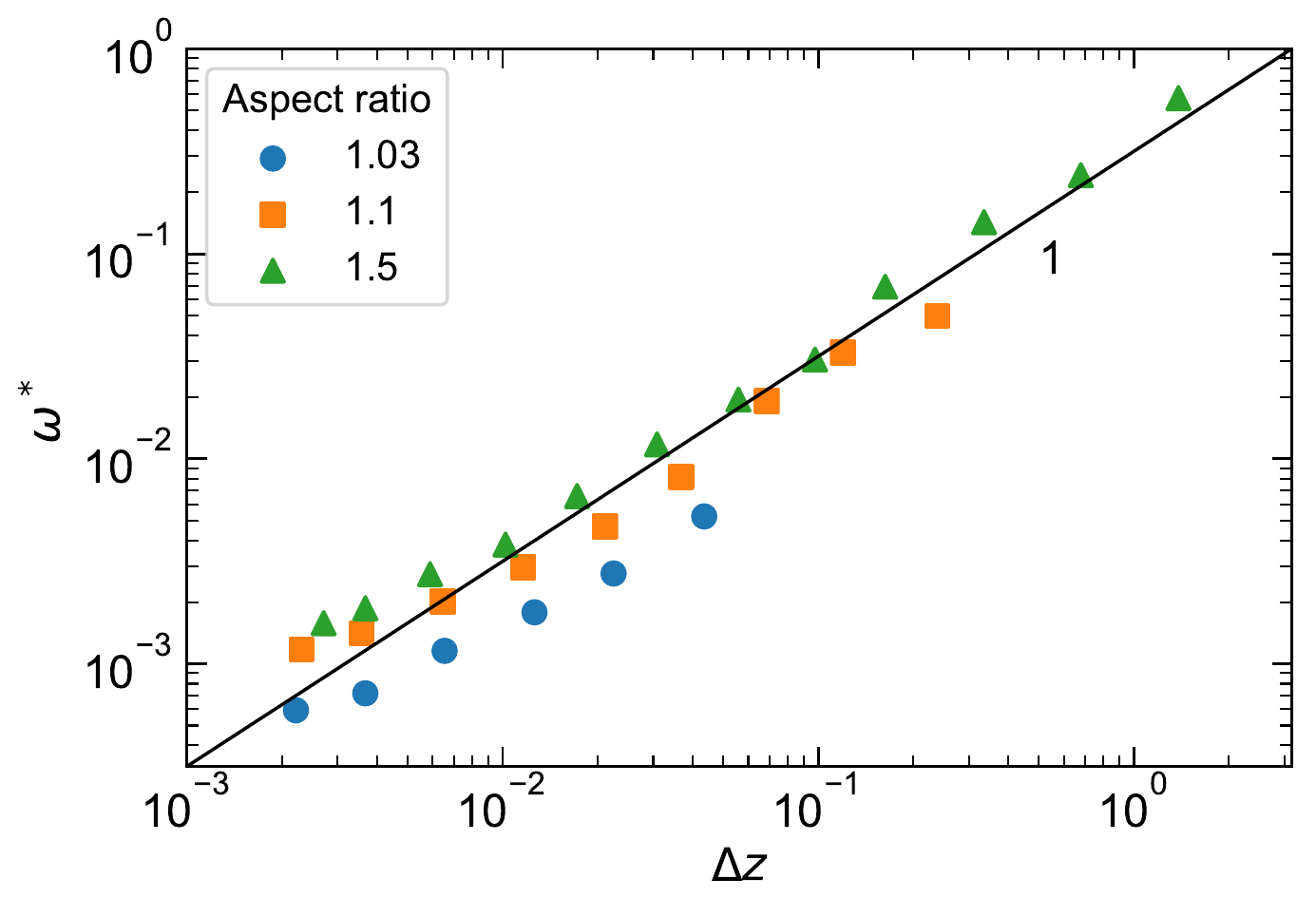}
 \caption{The frequency $\omega^{*}$ to characterize the onset of the rotational plateau in the vDOS.
We plot $\omega^{*}$ as a function of the excess contact number $\Delta z$ for different values of aspect ratio $\alpha$.
$\omega^{*}$ scales linearly with $\Delta z$, as $\omega^{*} \propto \Delta z$.
Additionally, as $\alpha$ becomes closer to unity, $\omega^{*}$ decreases.}
 \label{fig:omegastar-dz}
\end{figure}

We next focus on the excess density $\Delta\phi$ and aspect ratio $\alpha$ dependencies of $\omega^{*}$.
Figure~\ref{fig:omegastar-dz} plots $\omega^{*}$ as a function of the excess contact number $\Delta z$ for different values of aspect ratio $\alpha = \num{1.03},\ \num{1.1},\ \num{1.5}$~\footnote{Practically, the value of $\omega^{*}$ is determined as the frequency where $g\pqty{\omega}/\omega^{1/4}$ takes a maximum.}.
Remarkably, $\omega^{*}$ scales linearly with $\Delta z$, as $\omega^{*} \propto \Delta z$.
For spherical particles, the vDOS shows the characteristic plateau, and its onset frequency $\omega^{*}$ also scales linearly with $\Delta z$~\cite{Silbert_2005}.
Therefore, our result establishes the common scaling law for the onset of the ``rotational plateau'' of dimers and the plateau of spheres.
In addition, $\omega^{*}$ slightly decreases as the aspect ratio $\alpha$ goes to unity.
However, this tendency is very subtle, and the precise determination of the $\alpha$ dependence of $\omega^{*}$ requires the data at $\alpha$ to be much closer to unity.
This requires conducting simulations with a much larger system size, which are not available in this work.

\subsubsection{Real space displacements in the plateau modes}\label{sec:plateaumode}

\begin{figure}[tb]
 \centering
 \includegraphics[width=.99\columnwidth]{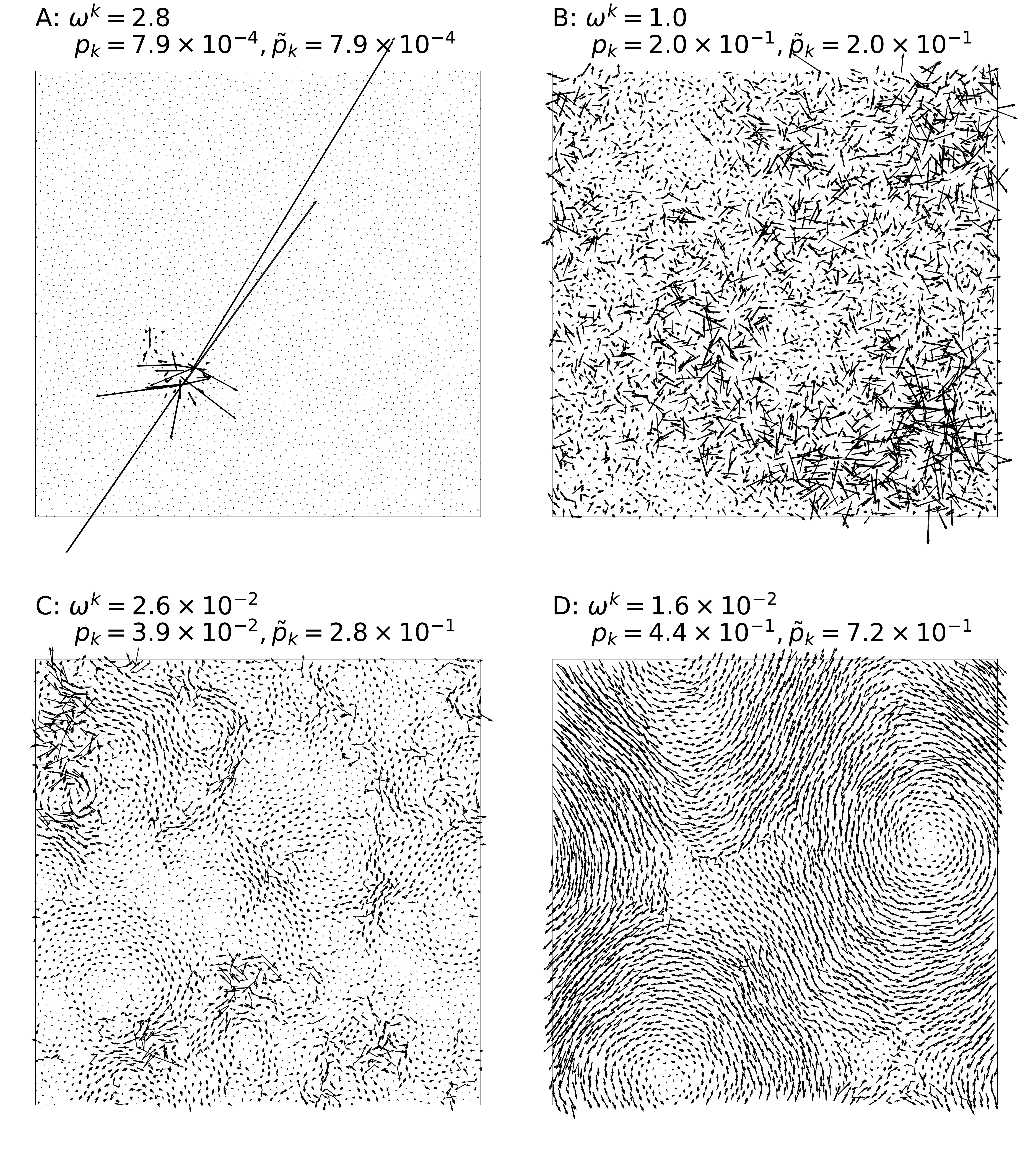}
 \caption{Visualization of vibrational eigenmodes for the case of aspect ratio $\alpha = \num{1.1}$.
The excess volume fraction is $\Delta\phi = \num{e-3}$, and the system size is $N = \num{3600}$.
(A) to (D) correspond to the frequency position of each visualized mode, which are indicated in the panel of $g\pqty{\omega}$ of Fig.~\ref{fig:DOS_Rk_PR_a1.1}: (A)~high-frequency edge, (B)~translational plateau, (C)~rotational plateau, and (D)~the low-frequency edge.
The values of frequencies and participation ratios of these modes are provided in the figures.}
 \label{fig:mode_visualize}
\end{figure}

In the previous subsections, we established that the vDOSs of the dimers with a small aspect ratio have the translational plateau and rotational plateau, which exhibit the critical behaviors near the jamming transition.
These two plateaus can be distinguished by the rotational DOFs, i.e., they are relevant in the rotational plateau but not in the translational plateau.
This result is true in the space of normal coordinates.
This subsection focuses on the real space displacement fields in the translational and rotational plateau modes.

We analyze the displacement fields of the vibrational modes in real space.
In normal coordinates, the vibrational eigenstates are expressed by the translational and rotational components, $\vb{e}^k_i = \pmqty{e^k_{i,x} & e^k_{i,y} & e^k_{i,\varphi}}$, for dimer $i$.
However, we can also express the vibrational eigenstates as displacement fields of monomers, which we denote by $\tilde{\vb{e}}^k_i = \pmqty{e^k_{\pqty{i,+1}, x} & e^k_{\pqty{i,+1}, y} & e^k_{\pqty{i,-1}, x} & e^k_{\pqty{i,-1}, y}}$ for dimer $i$.
These two descriptions can be transformed from each other by
\begin{align}
 \tilde{e}^k_{\pqty{i,\pm 1}, x} &= e^k_{i,x} \pm \frac{b_i}{2}\pqty{\alpha-1} \bqty{ \cos\pqty{ \frac{\varphi_i + e^k_{i,\varphi}}{\sqrt{I_i}} } - \cos\frac{\varphi_i}{\sqrt{I_i}} }, \label{transx}\\
 \tilde{e}^k_{\pqty{i,\pm 1}, y} &= e^k_{i,y} \pm \frac{b_i}{2}\pqty{\alpha-1} \bqty{ \sin\pqty{ \frac{\varphi_i + e^k_{i,\varphi}}{\sqrt{I_i}} } - \sin\frac{\varphi_i}{\sqrt{I_i}} }. \label{transy}
\end{align}
We note that $\vb{e}^k = \pmqty{\vb{e}^k_1 & \dots & \vb{e}^k_N}$ describes the vibrations in the phase space of $(x,y,\varphi)$, where translational and rotational motions are treated as different objects.
On the other hand, $\tilde{\vb{e}}^k = \pmqty{\tilde{\vb{e}}^k_1 & \dots & \tilde{\vb{e}}^k_N}$ describes the vibrations in the real space of $(x,y)$, where both translational and rotational motions are expressed as translational motions of monomers, namely, treated on equal footing.
We can define the participation ratio for the real space displacement fields as 
\begin{align}
 \tilde{p}_k = \frac{\pqty{\sum^{2N}_{i=1}\abs{\tilde{\vb{e}}^k_i}^2}^2}{2N\sum^{2N}_{i=1}\abs{\tilde{\vb{e}}^k_i}^4}.
\end{align}
We can also define the contribution of the rotational DOFs for the real space displacement fields as
\begin{align}
 \tilde{R}_k = \frac{\abs{\tilde{\vb{e}}^k - \tilde{\vb{e}}^k_\text{trans}}}{\abs{\tilde{\vb{e}}^k}}.
\end{align}
Here, $\tilde{\vb{e}}^k_\text{trans}$ is defined as
\begin{align}
 \tilde{e}^k_{\text{trans}\ \pqty{i,\pm 1}, x} &= e^k_{i,x},\\
 \tilde{e}^k_{\text{trans}\ \pqty{i,\pm 1}, y} &= e^k_{i,y},
\end{align}
which corresponds to the displacement fields of monomers that originated from only the translational DOFs in $\vb{e}^k$.

In Fig.~\ref{fig:mode_visualize}, we visualize the displacement fields, $\tilde{\vb{e}}^k$, in four representative eigenstates for the aspect ratio $\alpha = \num{1.1}$, whose frequencies are indicated in the top panel of Fig.~\ref{fig:DOS_Rk_PR_a1.1}.
The panel (A) shows a mode in the high-frequency edge.
This vibrational state is highly localized onto a few particles.
The panel (D) shows a vibrational state in the low frequency edge below the rotational plateau.
This vibrational state is more ordered and spatially extended, which exhibits a feature of an acoustic phonon.
These features are similar to those of the corresponding vibrational modes in sphere packings~\cite{Silbert_Liu_Nagel_2009}.

\begin{figure}[tb]
 \centering
 \includegraphics[width=.99\columnwidth]{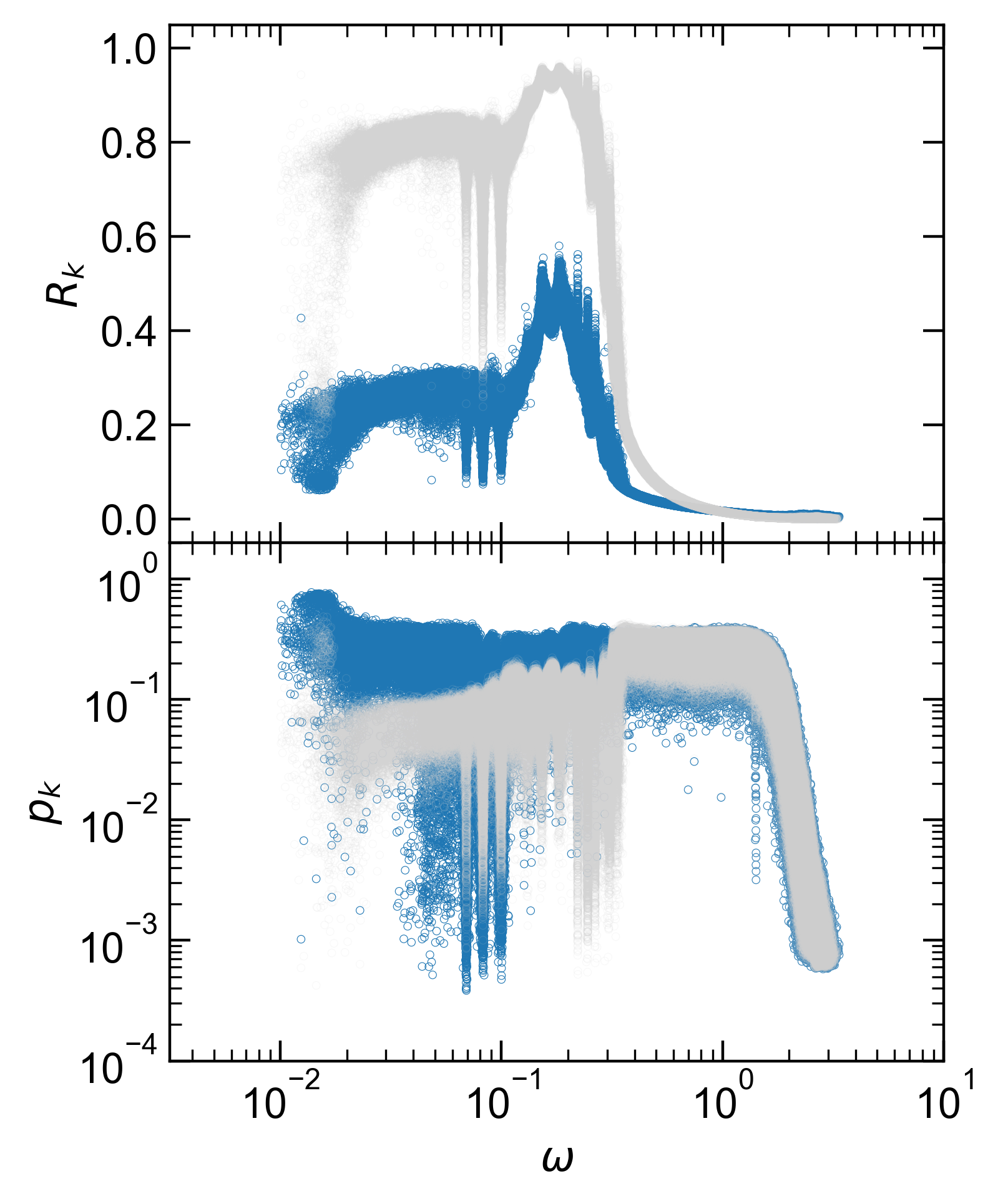}
 \caption{Analysis on real space displacements in the vibrational eigenmodes for the aspect ratio $\alpha = \num{1.1}$.
We plot the contribution of rotational DOFs $\tilde{R}_k$ and the participation ratios $\tilde{p}_k$.
The excess volume fraction is $\Delta\phi = \num{e-3}$, and the system size is $N = \num{3600}$.
For comparison, we also plot the values of $R_k$ and $p_k$ based on the normal coordinates, which are presented in Fig.~\ref{fig:DOS_Rk_PR_a1.1}.}
 \label{fig:DOS_tilde_a1.1}
\end{figure}

Now, we discuss the features of translational and rotational plateau modes.
The panel (B) shows a translational plateau mode, while the panel (C) shows a rotational plateau mode.
In both cases, the vibrational states are disordered and spatially extended.
These features are very similar to those of the plateau modes in sphere packings~\cite{Silbert_Liu_Nagel_2009}.
We note that the rotational plateau mode (C) appears to be more ``swirly'' compared to the translational plateau mode (B).
However, this difference cannot be attributed to the rotational nature of the modes because sphere packings already show similar swirly motions in the vibrational modes in the lower-frequency part of the plateau~\cite{van_Hecke_2010,Silbert_Liu_Nagel_2009}.
Apart from the swirly motions, the visualization of real space displacements does not indicate any strong difference between these two types of modes.
In contrast, the values of $R_k$ and $p_k$ indicate large differences between these two modes, as shown in Fig.~\ref{fig:DOS_Rk_PR_a1.1}.
We discuss this point below.

First, the displacement field of the rotational plateau mode (C) does not appear to be very ``rotational'', regardless of high values of $R_k \approx \num{0.8}$.
In the displacement fields, the purely rotational motion of dimers appears as the pair of arrows (rooted at two monomers) with the opposite directions.
However in the panel (C), such pairs of arrows cannot be found.
This visual inspection is supported by the values of $\tilde{R}_k$ shown in Fig.~\ref{fig:DOS_tilde_a1.1}: $\tilde{R}_k$ indicates much lower values of $\tilde{R}_k \approx \num{0.2}$ than those of $R_k \approx \num{0.8}$ for the rotational plateau modes.
The large difference between $R_k$ and $\tilde{R}_k$ can be understood when we go back to the transformation formula Eqs.~\eqref{transx} and~\eqref{transy}; the factor of $\pqty{\alpha - 1}$ suppresses the impact of the rotation in the displacements in the real space.
This situation is analogous to the rotation of a rod.
When the rod is very short, a large rotation does not cause large displacements of its edges.
Therefore, these results show that the translational motions of dimers are more dominant than the rotational motions of dimers for the real space displacements, even for the rotational plateau modes.

In addition, the displacement field of the rotational plateau mode (C) does not appear to be very localized, regardless of low values of $p_k \lesssim \num{0.1}$.
This visual inspection is again supported by the values of $\tilde{p}_k$ shown in Fig.~\ref{fig:DOS_tilde_a1.1}: $\tilde{p}_k$ indicates the higher values of $\tilde{p}_k \approx \num{0.3}$ than those of $p_k \lesssim \num{0.1}$ for the rotational plateau modes, which are almost the same as those of the translational plateau modes.
This observation suggests that the vibrational localization takes place only in the rotational DOFs that do not strongly affect the real space displacements.

In summary, these analyses reveal that the real space displacements for both the translational and rotational plateau modes are disordered and spatially extended.
Remarkably, the displacement field of the rotational plateau mode is still dominated by the translational motions of dimers, and the rotational motions only introduce slight modifications to it, as evidenced by the low values of $\tilde{R}_k \approx \num{0.2}$.
We emphasize, however, that the distinction between the translational and rotational plateau modes is still clear and definite.
For translational plateau modes, the rotational motions are fully frozen and never introduce even small modifications to the displacement field, as evidenced from nearly zero values of $\tilde{R}_k \approx \num{0}$.

\section{Conclusion}\label{sec:conclusion}
In this work, we study critical behaviors of vibrational properties near the jamming transition for dimer packings in two dimensions.
Our dimer particle is modeled by connecting two harmonic disks.
However, we revealed that this usual modeling of dimers encounters problems that involve the unusual contacts between dimers, i.e., the double contact and the cusp contact.
We then clarified that the unusual contacts appear only at high packing fractions and that the model is free from this problem near the jamming transition.
Our main results are obtained from the density region without this problem.

We first studied the contact number of dimers near the jamming transition.
We found that the contact number at the jamming transition, $z_J$, is smaller than the naive isostatic contact number of dimers.
This apparent shortage of contact number originated from the presence of ``rotational rattlers'', in which the dimer has three contacts or more, but these contacts are concentrated in one monomer of the dimer so that the dimer still has a free rotational motion.
Therefore, we corrected this effect of rotational rattlers on the isostatic contact number and defined the excess contact number $\Delta z$ as that measured from this corrected isostatic contact number.
We then established that $\Delta z$ defined in this way obeys the same critical law and finite size scaling law as those of sphere packings; namely, it follows the square root law of $\Delta z \propto \Delta\phi^{1/2}$ for large systems, as well as the finite size scaling law of $N \Delta z = G \pqty{\Delta\phi^{1/2}N}$, where $G\pqty{x} \sim 1$ for small $x$ and $G\pqty{x} \sim x$ for large $x$.

We next studied the vibrational properties of dimers near the jamming transition.
The vDOS of dimers with a large aspect ratio is composed of a high-frequency edge, the plateau in the intermediate frequency, and the acoustic phonon in the low-frequency edge, which are very similar to the features of vDOS of spheres.
We note, however, that the vibrational states in the intermediate plateau have a rotational nature.
Interestingly, for the case of the small aspect ratio, the plateau at the intermediate frequency is separated into two plateaus by a peak.
The vibrational states of the high- and low-frequency plateaus are dominated by the translational and rotational DOFs, respectively, which are thus called ``translational plateau'' and ``rotational plateau''.
We also call the peak that separates the two plateaus the rotational peak.
However, we remark that the displacement fields in the real space are still dominated by the translational DOFs even for the modes in the rotational plateau.

In addition, we defined two frequencies to characterize the above features of vDOS: $\omega_R$ for the frequency of the rotational peak and $\omega^{*}$ for the onset of the rotational plateau.
The frequency $\omega_R$ to characterize the rotational peak is independent of the packing fraction but depends on the aspect ratio as $\omega_R \propto \pqty{\alpha - 1}$, which is reasonable as the fundamental frequency scale of rotational vibrations in this model.
On the other hand, the onset of the rotational plateau $\omega^{*}$ depends on the packing fraction; it scales as $\omega^{*} \propto \Delta z$, which is the same scaling law as that of the plateau in spheres.
These results establish a fundamental view on the vibrational properties of dimers near the jamming transition.
In this work, we restricted ourselves to the vibrational properties of the unstressed system.
Of course, these findings should be supplemented with studies of original, stressed systems.
We are now working on this direction.

As a final remark, we would compare the vibrational properties of dimers with those of ellipsoids~\cite{Schreck_2012,Brito_2018}.
The characteristic frequencies of the rotational bands scale linearly with $\alpha - 1$ in ellipsoids, which is the same as the scaling behavior of $\omega_R$ in dimers.
On the other hand, ellipsoids have the quartic vibrational modes in the lowest-frequency region, which are totally absent in dimers.
Instead, dimers have the extended rotational plateau in the vDOS, which does not appear in the vDOS of ellipsoids.
We tentatively attribute these differences in vibrational properties to the difference between ``hypostaticity'' of ellipsoids and ``isostaticity'' of dimers.
However, for a deeper understanding of the rotational plateau modes in dimer packings, we need further numerical and theoretical studies.

\begin{acknowledgments}
We thank Harukuni~Ikeda and Daniele~Coslovich for useful discussions and suggestions.
This work was supported by JSPS KAKENHI Grant Numbers 17K14369, 17H04853, 16H04034, 18H05225, and 19K14670.
This work was also partially supported by the Asahi Glass Foundation.
The computations were partially performed using the Research Center for Computational Science, Okazaki, Japan.
\end{acknowledgments}

\appendix
\section{Moment of inertia of a dimer}\label{app:moment_of_inertia}
In this appendix, we provide the formulation of the moment of inertia of a dimer $i$, $I_i$.

\begin{figure}[tb]
 \centering
 \includegraphics[width=.99\columnwidth]{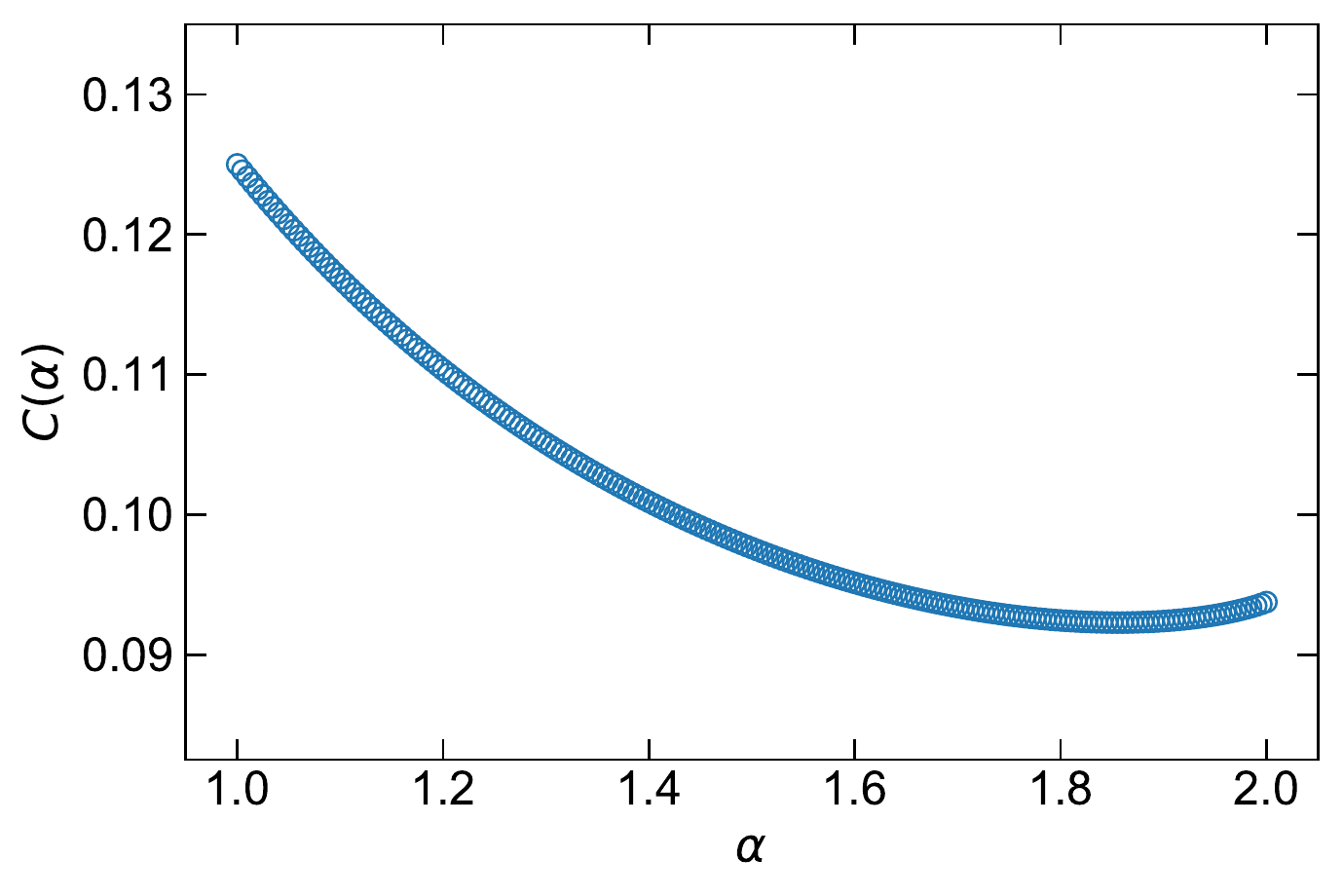}
 \caption{$\alpha$ dependency of $C\pqty{\alpha}$, shown in Eq.~\eqref{C_alpha}.}
 \label{fig:C_alpha}
\end{figure}

We set a coordinate system, $\pqty{x,y}$, on the dimer particle that is illustrated in Fig.~\ref{fig:dimer_shape}.
We set the origin of this coordinate system at the center of the dimer, and $x$ and $y$ axes along the major and the minor axes of the dimer, respectively.
Then, the moment of inertia of the dimer $i$, $I_i$, is formulated as
\begin{align}
 I_i &= \int^{a_i/2}_{-a_i/2}\dd{x}\int^{y\pqty{x}}_{-y\pqty{x}}\dd{y}\rho_i\pqty{x^2 + y^2} \\
 &= \rho_i\int^{a_i/2}_{-a_i/2}\dd{x} 2x^2y\pqty{x} + \rho_i\int^{a_i/2}_{-a_i/2}\dd{x} \frac{2}{3}y\pqty{x}^3, \label{inertia1}
\end{align}
where $\rho_i$ is an areal density of the 2D dimer $i$ with:
\begin{align}
 \rho_i = \frac{m}{ \frac{{a_i}^2}{2\alpha^2} \bqty{\pi - \cos^{-1}\pqty{\alpha-1} + \pqty{\alpha-1}\sqrt{\alpha\pqty{2-\alpha}}} },
\end{align}
and $y\pqty{x}$ is
\begin{align}
 y\pqty{x} = \begin{cases}
             &\sqrt{\frac{b^2}{4} - \pqty{x-\frac{a-b}{2}}^2}, \quad x \geq 0,\\
             &\sqrt{\frac{b^2}{4} - \pqty{x+\frac{a-b}{2}}^2}, \quad x < 0.
            \end{cases}
\end{align}
By performing the integration in Eq.~\eqref{inertia1}, we obtain the moment of inertia $I_i$ as
\begin{align}
 I_i = {a_i}^2 C\pqty{\alpha},
\end{align}
where $C\pqty{\alpha}$ is a function of only the aspect ratio $\alpha$:
\begin{widetext}
\begin{align}
 C\pqty{\alpha} = \frac{2\alpha^2}{\pi - \cos^{-1}\pqty{\alpha-1} + \pqty{\alpha-1}\sqrt{\alpha\pqty{2-\alpha}}} \left\{ \int_{-1/2}^0 \dd{x} \sqrt{\frac{1}{4\alpha^2} - \pqty{x + \frac{\alpha-1}{2\alpha}}^2} \bqty{2x^2 + \frac{2}{3}\pqty{\frac{1}{4\alpha^2} - \pqty{x + \frac{\alpha-1}{2\alpha}}^2}} \right. \nonumber \\
\left. + \int_0^{1/2}  \dd{x} \sqrt{\frac{1}{4\alpha^2} - \pqty{x - \frac{\alpha-1}{2\alpha}}^2} \bqty{2x^2 + \frac{2}{3}\pqty{\frac{1}{4\alpha^2} - \pqty{x - \frac{\alpha-1}{2\alpha}}^2}} \right\} . \label{C_alpha}
\end{align}
\end{widetext}
The behavior of $C\pqty{\alpha}$ is shown in Fig.~\ref{fig:C_alpha}.

\section{Dynamical matrix}\label{app:dynmat}
Here, we provide the explicit formulations of the elements of dynamical matrix, $\mathcal{M}_{kl}$, defined in Eq.~\eqref{dynmat} ($k,l = 1, \dots, d_fN$).

The distance between the monomer $k_i$ of the dimer $i$ and the monomer $k_j$ of the dimer $j$ is formulated as
\begin{align}
 r_{k_ik_j} &= \sqrt{X_{k_ik_j}^2 + Y_{k_ik_j}^2},\\
 X_{k_ik_j} &= x_j - x_i + \frac{\alpha-1}{2}\bqty{ b_j n_j \cos \frac{\varphi_j}{\sqrt{I_j}} - b_i n_i \cos \frac{\varphi_i}{\sqrt{I_i}} },\\
 Y_{k_ik_j} &= y_j - y_i + \frac{\alpha-1}{2}\bqty{ b_j n_j \sin \frac{\varphi_j}{\sqrt{I_j}} - b_i n_i \sin \frac{\varphi_i}{\sqrt{I_i}} },
\end{align}
where $n_i$~($n_j$) takes values of $1$ or $-1$ to represent the monomer of the dimer $i$~($j$).
Additionally, the force $\drv{i}{j}$ and the stiffness $\ddrv{i}{j}$ acting between these two monomers (which correspond to the first and second derivatives of the potential, Eq.~(\ref{potential}), respectively) are
\begin{align}
 \drv{i}{j}  &\equiv \frac{\epsilon}{\sigma_{ij}}\pqty{1-\frac{r_{k_ik_j}}{\sigma_{ij}}} H\pqty{1-\frac{r_{k_ik_j}}{\sigma_{ij}}},\\
 \ddrv{i}{j} &\equiv \frac{\epsilon}{{\sigma_{ij}}^2} H\pqty{1-\frac{r_{k_ik_j}}{\sigma_{ij}}},
\end{align}
where we recall that $H\pqty{x}$ is the Heaviside step function: $H\pqty{x} = 1$ for $x \geq 0$ and $H\pqty{x} = 0$ for $x < 0$.
Then, we can formulate the elements of dynamical matrix, $\mathcal{M}_{kl}$, as follows.
Note that we use variables $\xi_i$ to designate elements of the dynamical matrix ($\xi = x, y, \varphi$) in the following.
In the sense of the definition in Eq.~\eqref{dynmat}, this should be converted as $k = d_f\pqty{i-1} + n_\xi$, where $n_\xi = 1$ for $\xi = x$, $n_\xi = 2$ for $\xi = y$, $n_\xi = 3$ for $\xi = \varphi$.
We recall here that $k_i$ is a set of integers composed of $\pqty{i, n_i}$, and represents a monomer of dimer $i$.

\begin{widetext}
The elements for $i=j$ are
\begin{align}
 \mathcal{M}_{x_a x_a} &= \sum^N_{i \neq a} \sum_{n_a n_i} \pqty{\ddrv{a}{i}+\frac{\drv{a}{i}}{r_{k_ak_i}}} \frac{X_{k_ak_i}^2}{r_{k_ak_i}^2} - \frac{\drv{a}{i}}{r_{k_ak_i}},\\
 \mathcal{M}_{y_a y_a} &= \sum^N_{i \neq a} \sum_{n_a n_i} \pqty{\ddrv{a}{i}+\frac{\drv{a}{i}}{r_{k_ak_i}}} \frac{Y_{k_ak_i}^2}{r_{k_ak_i}^2} - \frac{\drv{a}{i}}{r_{k_ak_i}},\\
 \mathcal{M}_{\varphi_a \varphi_a} &= \sum^N_{i \neq a} \sum_{n_a n_i} \pqty{\ddrv{a}{i}+\frac{\drv{a}{i}}{r_{k_ak_i}}} \pqty{\frac{b_an_a\pqty{\alpha-1}}{2r_{k_ak_i}\sqrt{I_a}}}^2 \pqty{-X_{k_ik_a}\sin\frac{\varphi_a}{\sqrt{I_a}} + Y_{k_ak_i}\cos\frac{\varphi_a}{\sqrt{I_a}}}^2 \nonumber \\
&- \drv{a}{i} \frac{b_an_a\pqty{\alpha-1}}{2r_{k_ak_i}\sqrt{I_a}}\pqty{\frac{b_an_a\pqty{\alpha-1}}{2\sqrt{I_a}} - \frac{X_{k_ak_i}}{\sqrt{I_a}}\cos\frac{\varphi_a}{\sqrt{I_a}} - \frac{Y_{k_ak_i}}{\sqrt{I_a}}\sin\frac{\varphi_a}{\sqrt{I_a}}},\\
 \mathcal{M}_{x_a y_a} &= \sum^N_{i \neq a} \sum_{n_a n_i} \pqty{\ddrv{a}{i}+\frac{\drv{a}{i}}{r_{k_ak_i}}} \frac{X_{k_ak_i}Y_{k_ak_i}}{r_{k_ak_i}^2},\\
 \mathcal{M}_{x_a \varphi_a} &= \sum^N_{i \neq a} \sum_{n_a n_i} -\pqty{\ddrv{a}{i}+\frac{\drv{a}{i}}{r_{k_ak_i}}} \frac{X_{k_ak_i}}{r_{k_ak_i}} \frac{b_an_a\pqty{\alpha-1}}{2r_{k_ak_i}\sqrt{I_a}} \pqty{X_{k_ak_i}\sin\frac{\varphi_a}{\sqrt{I_a}} - Y_{k_ak_i}\cos\frac{\varphi_a}{\sqrt{I_a}}} + \drv{a}{i} \frac{b_an_a\pqty{\alpha-1}}{2r_{k_ak_i}\sqrt{I_a}}\sin\frac{\varphi_a}{\sqrt{I_a}},\\
 \mathcal{M}_{y_a \varphi_a} &= \sum^N_{i \neq a} \sum_{n_a n_i} -\pqty{\ddrv{a}{i}+\frac{\drv{a}{i}}{r_{k_ak_i}}} \frac{Y_{k_ak_i}}{r_{k_ak_i}} \frac{b_an_a\pqty{\alpha-1}}{2r_{k_ak_i}\sqrt{I_a}} \pqty{X_{k_ak_i}\sin\frac{\varphi_a}{\sqrt{I_a}} - Y_{k_ak_i}\cos\frac{\varphi_a}{\sqrt{I_a}}} - \drv{a}{i} \frac{b_an_a\pqty{\alpha-1}}{2r_{k_ak_i}\sqrt{I_a}}\cos\frac{\varphi_a}{\sqrt{I_a}},
\end{align}
and the elements for $i \neq j$ are
\begin{align}
 \mathcal{M}_{x_i x_j} &= \sum_{n_i n_j} -\pqty{\ddrv{i}{j}+\frac{\drv{i}{j}}{r_{k_ik_j}}}\frac{X_{k_ik_j}^2}{r_{k_ik_j}^2} + \frac{\drv{i}{j}}{r_{k_ik_j}},\\
 \mathcal{M}_{y_i y_j} &= \sum_{n_i n_j} -\pqty{\ddrv{i}{j}+\frac{\drv{i}{j}}{r_{k_ik_j}}}\frac{Y_{k_ik_j}^2}{r_{k_ik_j}^2} + \frac{\drv{i}{j}}{r_{k_ik_j}},\\
 \mathcal{M}_{\varphi_i \varphi_j} &= \sum_{n_i n_j} - \pqty{\ddrv{i}{j}+\frac{\drv{i}{j}}{r_{k_ik_j}}} \frac{b_in_i b_jn_j \pqty{\alpha-1}^2}{4r_{k_ik_j}^2\sqrt{I_i}\sqrt{I_j}} \pqty{X_{k_ik_j}\sin\frac{\varphi_i}{\sqrt{I_i}}-Y_{k_ik_j}\cos\frac{\varphi_i}{\sqrt{I_i}}}\pqty{X_{k_ik_j}\sin\frac{\varphi_j}{\sqrt{I_j}}-Y_{k_ik_j}\cos\frac{\varphi_j}{\sqrt{I_j}}} \nonumber \\
 &+ \drv{i}{j}\frac{b_in_i b_jn_j \pqty{\alpha-1}^2}{4 r_{k_ik_j} \sqrt{I_i}\sqrt{I_j}}\cos\pqty{\frac{\varphi_i}{\sqrt{I_i}} - \frac{\varphi_j}{\sqrt{I_j}}}, \label{Mthetatheta}\\
 \mathcal{M}_{x_i \varphi_j} &= \sum_{n_i n_j} \pqty{\ddrv{i}{j}+\frac{\drv{i}{j}}{r_{k_ik_j}}} \frac{X_{k_ik_j}}{r_{k_ik_j}} \frac{b_jn_j\pqty{\alpha-1}}{2r_{k_ik_i}\sqrt{I_j}} \pqty{X_{k_ik_j}\sin\frac{\varphi_j}{\sqrt{I_j}}-Y_{k_ik_j}\cos\frac{\varphi_j}{\sqrt{I_j}}} - \drv{i}{j} \frac{b_jn_j \pqty{\alpha-1}}{2r_{k_ik_j}\sqrt{I_j}} \sin\frac{\varphi_j}{\sqrt{I_j}},\\
 \mathcal{M}_{\varphi_i x_j} &= \sum_{n_i n_j} \pqty{\ddrv{i}{j}+\frac{\drv{i}{j}}{r_{k_ik_j}}} \frac{X_{k_ik_j}}{r_{k_ik_j}} \frac{b_in_i\pqty{\alpha-1}}{2r_{k_ik_i}\sqrt{I_i}} \pqty{X_{k_ik_j}\sin\frac{\varphi_i}{\sqrt{I_i}}-Y_{k_ik_j}\cos\frac{\varphi_i}{\sqrt{I_i}}} - \drv{i}{j} \frac{b_in_i \pqty{\alpha-1}}{2r_{k_ik_j}\sqrt{I_i}} \sin\frac{\varphi_i}{\sqrt{I_i}},\\
 \mathcal{M}_{y_i \varphi_j} &= \sum_{n_i n_j} \pqty{\ddrv{i}{j}+\frac{\drv{i}{j}}{r_{k_ik_j}}} \frac{Y_{k_ik_j}}{r_{k_ik_j}} \frac{b_jn_j\pqty{\alpha-1}}{2r_{k_ik_j}\sqrt{I_j}} \pqty{X_{k_ik_j}\sin\frac{\varphi_j}{\sqrt{I_j}}-Y_{k_ik_j}\cos\frac{\varphi_j}{\sqrt{I_j}}} + \drv{i}{j} \frac{b_jn_j \pqty{\alpha-1}}{2r_{k_ik_j}\sqrt{I_j}} \cos\frac{\varphi_j}{\sqrt{I_j}},\\
 \mathcal{M}_{y_i \varphi_j} &= \sum_{n_i n_j} \pqty{\ddrv{i}{j}+\frac{\drv{i}{j}}{r_{k_ik_j}}} \frac{Y_{k_ik_j}}{r_{k_ik_j}} \frac{b_in_i\pqty{\alpha-1}}{2r_{k_ik_j}\sqrt{I_i}} \pqty{X_{k_ik_j}\sin\frac{\varphi_i}{\sqrt{I_i}}-Y_{k_ik_j}\cos\frac{\varphi_i}{\sqrt{I_i}}} + \drv{i}{j} \frac{b_in_i \pqty{\alpha-1}}{2r_{k_ik_j}\sqrt{I_i}} \cos\frac{\varphi_i}{\sqrt{I_i}},\\
 \mathcal{M}_{x_iy_j} &= \sum_{n_i n_j} -\pqty{\ddrv{i}{j} + \frac{\drv{i}{j}}{r_{k_ik_j}}}\frac{X_{k_ik_j}Y_{k_ik_j}}{r_{k_ik_j}^2},\\
 \mathcal{M}_{y_ix_j} &= \sum_{n_i n_j} -\pqty{\ddrv{i}{j} + \frac{\drv{i}{j}}{r_{k_ik_j}}}\frac{X_{k_ik_j}Y_{k_ik_j}}{r_{k_ik_j}^2}.
\end{align}
\end{widetext}

\bibliography{dimer.bib}
\end{document}